\documentclass[%
 aip,amssymb,
 reprint,
]{revtex4-1}

\usepackage{graphicx}
\usepackage{dcolumn}
\usepackage{bm}

\usepackage[utf8]{inputenc}
\usepackage[T1]{fontenc}
\usepackage{mathptmx}
\usepackage{etoolbox}
\usepackage{float}
\usepackage{color}
\usepackage{tikz}
\usepackage{amsmath}
\DeclareUnicodeCharacter{2212}{-}
\DeclareUnicodeCharacter{0301}{-}
\makeatletter
\def\@email#1#2{%
 \endgroup
 \patchcmd{\titleblock@produce}
  {\frontmatter@RRAPformat}
  {\frontmatter@RRAPformat{\produce@RRAP{*#1\href{mailto:#2}{#2}}}\frontmatter@RRAPformat}
  {}{}
}%
\makeatother
\begin{document}

\preprint{AIP/123-QED}

\title{Heterogeneous ice nucleation with TIP4P/Ice:
the role of lattice mismatch}

\author{M. Camarillo}
\affiliation{Departamento de Química Física, Facultad de Ciencias Químicas, Universidad Complutense de Madrid, 28040 Madrid, Spain}
\affiliation{Departamento de Ingeniería Química Industrial y del Medio Ambiente, Escuela T\'{e}cnica Superior de Ingenieros Industriales, Universidad Polit\'{e}cnica de Madrid, 28006, Madrid, Spain}

\author{J. Oller-Iscar}
\affiliation{Departamento de Ingeniería Química Industrial y del Medio Ambiente, Escuela T\'{e}cnica Superior de Ingenieros Industriales, Universidad Polit\'{e}cnica de Madrid, 28006, Madrid, Spain}

\author{M.M. Conde}
\affiliation{Departamento de Ingeniería Química Industrial y del Medio Ambiente, Escuela T\'{e}cnica Superior de Ingenieros Industriales, Universidad Polit\'{e}cnica de Madrid, 28006, Madrid, Spain}

\author{J. Ramírez}
\affiliation{Departamento de Ingeniería Química Industrial y del Medio Ambiente, Escuela T\'{e}cnica Superior de Ingenieros Industriales, Universidad Polit\'{e}cnica de Madrid, 28006, Madrid, Spain}

\author{S.Blazquez}
\affiliation{Departamento de Química Física, Facultad de Ciencias Químicas, Universidad Complutense de Madrid, 28040 Madrid, Spain}

\author{E. Sanz*}
\affiliation{Departamento de Química Física, Facultad de Ciencias Químicas, Universidad Complutense de Madrid, 28040 Madrid, Spain}
\email{esa01@ucm.es}

\date{\today}

\begin{abstract}

Ice formation, a crucial phase transition for life on Earth, mainly occurs through heterogeneous nucleation on a solid substrate. Despite the importance of this phenomenon, the correlation between the structure of the substrate and its ability to induce ice nucleation remains poorly understood.
In this work, we use molecular dynamics simulations with an atomistic model of water to investigate how the lattice mismatch affects the ability of the 
substrate to nucleate ice. To unambiguously quantify the mismatch, we use substrates with a stretched 
ice structure.
We put the substrate in contact with supercooled water and determine the temperature at which ice spontaneously nucleates during the course of the simulations. We find that the nucleation temperature decreases by $\sim$ \textcolor{black}{5-6} K for each percent by which the substrate lattice is stretched relative to that of ice. This trend is similar
to that previously obtained with a coarse-grained 
water model [Camarillo et al., J. Chem. Phys. 160, 134505 (2024)].
\textcolor{black}{The substrate induces a fluctuating network of subcritical ice-like clusters, from which a spherical-cap-shaped supercritical nucleus eventually emerges stochastically.}
Using Classical Nucleation Theory, combine previously reported homogeneous nucleation rates with heterogeneous nucleation rates computed in this work to estimate key heterogeneous nucleation parameters, including the nucleation barrier, the kinetic pre-factor, and the contact angle. The latter is consistent with a microscopic visualization of nucleation trajectories, supporting the applicability of Classical Nucleation Theory to describe a microscopic phenomenon such as heterogeneous ice nucleation.

\end{abstract}

\maketitle

\section{\label{sec:intro}Introduction}

Ice nucleation plays a central role in a wide range of natural and technological processes, including cloud formation, precipitation, climate regulation, cryopreservation, and icing phenomena \cite{pruppacher1998microphysics,cantrell2005production}. In most realistic environments, ice forms via heterogeneous nucleation, where the presence of a foreign substrate lowers the free-energy barrier for the formation of a critical ice nucleus \cite{hoose2012heterogeneous}. Understanding the microscopic mechanisms that govern heterogeneous ice nucleation is therefore essential for accurately describing and predicting freezing processes \cite{sosso2016crystal}.

In nature, ice-nucleating substrates exhibit an extraordinary diversity, ranging from mineral dust and soot particles to biological macromolecules such as proteins and bacteria \cite{pruppacher1998microphysics,cantrell2005production,hoose2012heterogeneous}. This diversity highlights the intrinsic complexity of heterogeneous ice nucleation, which depends sensitively on multiple factors, including the structural properties of the substrate, the nature and strength of substrate–water interactions, and the interfacial arrangement of water molecules. Disentangling the relative importance of these factors remains a major challenge.

Molecular dynamics simulations have emerged as a powerful tool to investigate heterogeneous ice nucleation at the molecular scale, providing direct access to structural, dynamical, and thermodynamic information that is often inaccessible experimentally \cite{lupi2014heterogeneous,zielke2015molecular,pedevilla2016can}. However, even within simulations, the problem remains highly nontrivial due to the activated nature of nucleation and to the subtle interplay between different factors that promote or inhibit ice formation at interfaces \cite{fitzner2015many,glatz2018heterogeneous,cox2012non}.

Among the factors influencing heterogeneous ice nucleation, lattice similarity between the substrate and ice has long been proposed as a key determinant of nucleation efficiency \cite{turnbull1950kinetics,mithen2014epitaxial}. A close match between the substrate lattice and the basal or prismatic planes of ice can facilitate templating effects that stabilize ice-like structures at the interface \cite{zielke2015molecular,soni2022ice,reinhardt2014effects,fraux2014note}. At the same time, specific interactions between water molecules and the substrate can either enhance or suppress nucleation, depending on their strength, range, and directional character \cite{reinhardt2014effects,lupi2014does,fitzner2015many,glatz2018heterogeneous}. The interplay between lattice matching and interaction details complicates attempts to establish universal principles governing ice nucleation
\cite{glatz2018heterogeneous,fitzner2015many,reinhardt2014effects}.

In our previous work, we focused on isolating the role of lattice similarity by employing idealized substrates composed of water molecules arranged in crystalline lattices \cite{camarillo2024effect}. By systematically stretching or compressing ice-like lattices, we quantified the nucleating ability of the substrate as a function of the lattice deformation with respect to bulk ice. This approach allowed us to directly assess the effect of lattice mismatch while leaving aside the role of substrate–water interaction specificity. These studies were carried out using a coarse-grained water model \cite{molinero2009water}, enabling efficient exploration of parameter space, but lacking the 
accuracy of a realistic water model. 

In the present work, we extend this framework by employing a realistic \cite{abascal2005potential}, atomistic water model that captures the detailed nature of water–water interactions. This allows us to assess to what extent the loss of nucleation efficiency induced by lattice deformation depends on the specific form of the interaction potential. By comparing results obtained from coarse-grained and realistic water models, we evaluate the robustness of lattice-based descriptors of nucleation efficiency against changes in microscopic interaction details.

Our results show that both models predict a similar dependence of the heterogeneous ice nucleation temperature on the degree of lattice deformation of the substrate relative to ice. This finding suggests that lattice similarity plays a dominant and robust role in determining ice nucleation efficiency, while the detailed form of the interaction potential primarily modulates, rather than controls, the overall trend. 
\textcolor{black}{Moreover, we identify a nucleation mechanism in which the nucleus
emerges from a fluctuating network of sub-critical nuclei. In addition, we find that the contact angle of such a
nucleus is well described by Classical Nucleation Theory.}
These insights contribute to a more unified understanding of heterogeneous ice nucleation and help clarify the relative roles of structural and energetic factors at ice-nucleating interfaces.

\section{\label{sec:method}Simulation Details}

This study of heterogeneous nucleation was performed using the TIP4P/Ice  model\cite{abascal2005potential}. 
Molecular dynamics simulations were run in the 
GROMACS\cite{hess2008gromacs} 2016 package in double precision.
We used a 2 fs time step to integrate the equations of motion using a Verlet algorithm\cite{spreiter1999classical}.
Simulations were carried out in the NVT ensemble. 
The Nose-Hoover thermostat\cite{martyna1992nose} with a 2 ps relaxation time was used to keep temperature
constant. 
\textcolor{black}{We have employed the NVT rather than the NVE ensemble to maintain precise
temperature control while keeping the system size computationally affordable.
Although latent heat is released during nucleation and subsequent crystal
growth, our previous comparison of NVE and NVT simulations showed that both
ensembles yield the same ice growth rates \cite{montero2019ice}, suggesting
that heat dissipation through the thermostat does not significantly affect the
crystallization dynamics.}
Electrostatic interactions were calculated using the Particle Mesh Ewald (PME)\cite{wheeler2002less} method with a real space cutoff of 10\,\text{\AA}, an interpolation order of 4, and a mesh spacing of 1\,\text{\AA}. Van der Waals forces, represented by a 
Lennard-Jones potential, were truncated by a simple cutoff at 10\,\text{\AA}, \textcolor{black}{with energy corrections applied to compensate for long range interactions}. The neighbour list was updated every integration step with a radius of 10\,\text{\AA}, and a grid-based search algorithm was used. The geometry of the water molecules was constrained using the LINCS algorithm with 6$^{th}$ order  and 4 iterations\cite{hess1997lincs,hess2008p}, ensuring structural stability throughout the simulation. The displacement of the center of mass of 
the simulation box was avoided at each step using a linear scheme.  

\begin{figure}[h] 
    \centering
       \includegraphics[width=0.35\textwidth]{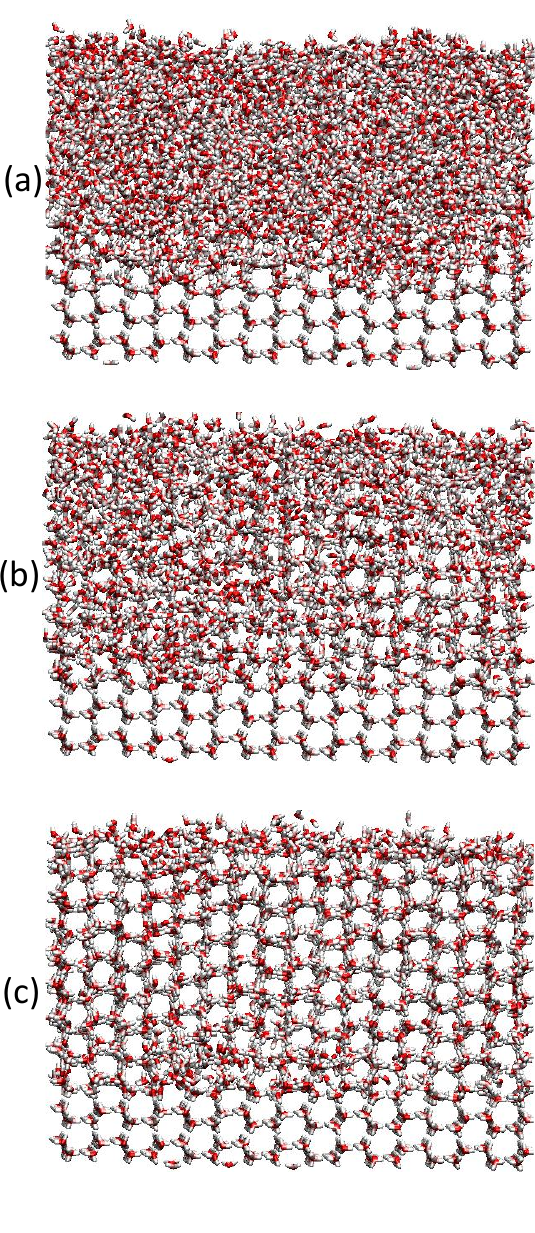}
    \caption{Snapshots of the 6\% mismatch ($\delta=$6) system at 244 K: (a) Starting configuration, (b) configuration with an emerging nucleus, (c) crystallized system. \textcolor{black}{The substrate is a stretched ice Ih lattice exposing the pI plane to the liquid. Panels (a), (b) and (c) correspond to snapshots taken at 0, 325 and 600 ns, 
    respectively, of
    the red trajectory shown in the top panel of Fig. \ref{fig:potencialenergy}.}}
    \label{fig:secuenciafotos}
\end{figure}

As in Ref. \cite{camarillo2024effect}, the substrate is built by stretching
in all 3 dimensions 
a
perfect ice \textcolor{black}{Ih} lattice at coexistence conditions
(1 bar and 270 K for TIP4P/Ice \cite{weiss2011kinetic,espinosa2016time}). 
The ice unit cell at coexistence conditions (with density 0.906 g/cm$^3$) is generated with GenIce software\cite{genice}. The unit cell is stretched
by a certain factor $f$ along all three
directions and replicated 1x8x8 along the
$x$, $y$ and $z$ coordinates to 
obtain the corresponding substrate.
The stretching percentage, $\delta$, is
given by $\delta=(f-1)\cdot100$.
The primary prismatic (pI) plane of the substrate is put in contact with a slab of liquid water. A free
interface \textcolor{black}{separating the liquid from the vapor phase} is left on the other
side of the liquid slab. \textcolor{black}{This geometry effectively fixes} pressure to 0 bar 
\textcolor{black}{(the vapor-liquid coexistence at the triple point is of the order of $10^{-4}$ bar for the TIP4P/Ice model \cite{vega2006vapor})}.
In Fig. \ref{fig:secuenciafotos}(a) we show a snapshot of a typical starting configuration. 
\textcolor{black}{The vapor phase (empty space) occupies about 40\% of the simulation box side perpendicular to the interface, $L_x$.}
Details of the simulated system sizes are found in Table \ref{tab:systems_parametres}.
Simulations are run 
under periodic 
boundary conditions (pbc) in all three directions. 
The positions and orientations of the substrate molecules are fixed during
the simulation (using the \emph{freeze} Gromacs option).

To have a visual representation of the nucleation 
pathway we identify the largest cluster of ice-like molecules
along the nucleation trajectory. 
For that purpose, we use the $\bar{q}_6$ order parameter \cite{lechner2008accurate}
to identify water molecules that have an ice-like local environment. 
\textcolor{black}{$\bar{q}_6$ is constructed from the spherical harmonics of degree $l=6$ evaluated along the directions connecting a molecule to its neighbors and subsequently averaged over the molecule and its neighboring molecule to reduce the effect of thermal fluctuations. Molecules in crystalline ice exhibit large values of $\bar{q}_6$ because of the strong orientational correlation between neighboring molecules. A threshold value, denoted by $\bar{q}_{6,t}$, is therefore used to classify molecules as ice-like ($\bar{q}_6>\bar{q}_{6,t}$) or liquid-like ($\bar{q}_6\leq\bar{q}_{6,t}$).}
For the specific threshold, $\bar{q}_{6,t}$, used to distinguish ice-like from liquid-like
water molecules, we employ $\bar{q}_{6,t}=0.364$ together with a neighbor cutoff
distance of 3.5~\AA{}, as in Ref.~\cite{espinosa2016time}. 
\textcolor{black}{Two ice-like molecules belong to
the same cluster if their mutual distance is $\leq$ 3.5~\AA. The threshold value of 0.364 was determined using the mislabeling criterion
introduced in Ref.~\cite{espinosa2016time}, which selects the threshold that equalizes the number of incorrectly classified molecules in the bulk ice and liquid phases~\cite{espinosa2016seeding}. In principle, the optimal threshold depends weakly on temperature, and the value of 0.364 was calibrated at 250~K. However, in the present work the largest ice cluster is used only to identify nucleation and subsequent crystal growth, rather than to determine thermodynamic or structural properties that depend sensitively on the exact cluster size. Consequently, using the same threshold for all temperatures considered here has a negligible effect on our results.}

\begin{table}[h!]
\centering
\begin{tabular}{c c c c c c c}
\hline
\hline
$\delta$ & $A/\mathrm{nm}^2$ & $N_\text{subs}$ & $N_\text{liq}$ & L$_x\,/\,\mathrm{nm}$ & L$_y\,/\,\mathrm{nm}$ & L$_z\,/\,\mathrm{nm}$ \\
\hline
4   & 46.49 & 1024 & 5120 & 8.82 & 6.15 & 7.56 \\
5   & 47.38 & 1024 & 5120 & 7.83 & 6.21 & 7.63 \\
5  & 47.38 & 1024 & 7168 &10.78 & 6.21 & 7.63 \\
6   & 48.28 & 1024 & 5120 & 7.33 & 6.27 & 7.70 \\
7   & 49.18 & 1024 & 5120 & 8.04 & 6.33 & 7.78 \\
\hline
\hline
\end{tabular}
\caption{Parameters describing the size of the different systems under study.
 $\delta$ is the percentage of elongation of the substrate lattice with respect
 to that of ice (thermalized at coexistence conditions with liquid water), $A$ is the substrate surface area, $N_\text{subs}$ and $N_\text{liq}$ correspond to the number of water molecules in the substrate and in the liquid, respectively; and L$_x$, L$_y$, L$_z$ represent the dimensions of the simulation box along each Cartesian axis, being the $x$ direction perpendicular
 to the substrate-liquid interface.}
\label{tab:systems_parametres}
\end{table}




\section{Results}
\subsection{Nucleation rate and nucleation temperature}

\begin{figure}[h] 
    \centering
    \includegraphics[width=0.48\textwidth]{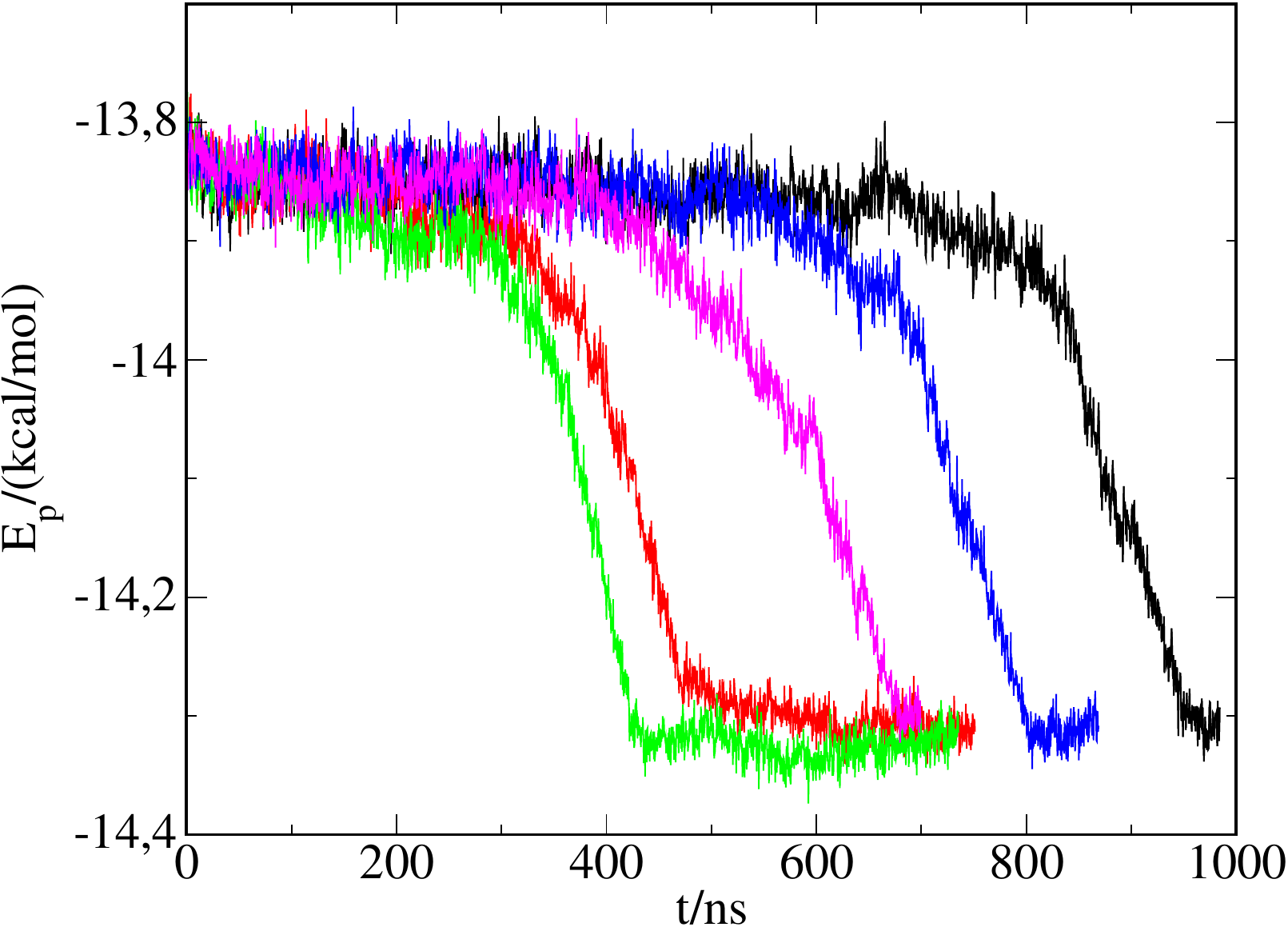} 
       \includegraphics[width=0.48\textwidth]{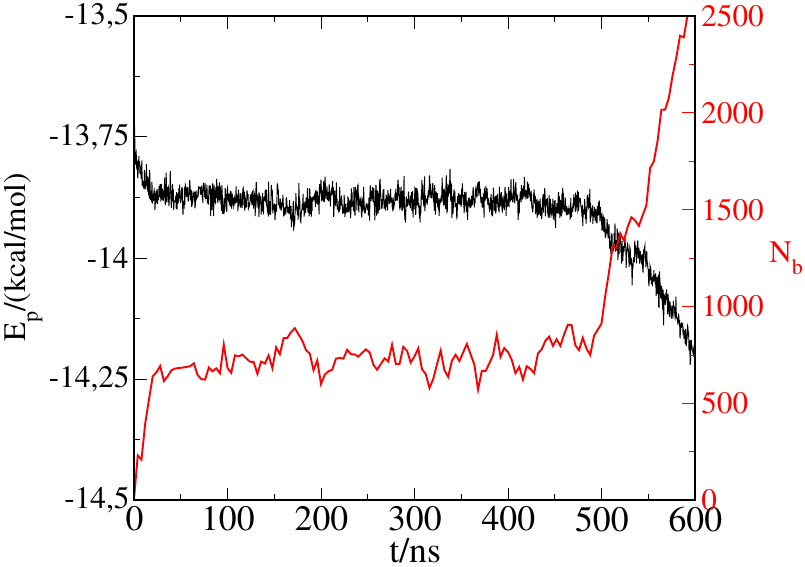}
    \caption{\textcolor{black}{Top:} Potential energy of the system versus time for different trajectories on an ice substrate stretched by 6 \% ($\delta=$6) at 244 K.
    \textcolor{black}{Bottom: energyenergy and number of molecules in the biggest ice cluster, $N_b$, versus time for a representative trajectory with a $\delta = 5$ substrate at 250 K.}}
    \label{fig:potencialenergy}
\end{figure}

We run long molecular dynamics simulations of supercooled water in contact with the substrate 
and wait for a nucleation event to occur after some induction period.
When an ice nucleus appears and grows, liquid water turns into ice and the potential energy shows a sudden drop, 
which enables us to identify the nucleation event. 
This phenomenology is illustrated in Fig. \ref{fig:potencialenergy}, where
we plot the potential energy versus time for 5 different trajectories
starting from the same configuration but differing in the seed
used to generate initial velocities. 
The potential energy drop takes place at different times
for different trajectories, which reflects the stochastic nature
of nucleation. 
The behavior 
illustrated in Fig. \ref{fig:potencialenergy}
is 
found only in a narrow temperature window. 
In particular, the example of Fig. \ref{fig:potencialenergy}
corresponds to 244 K and a substrate of $\delta=$ 6. 
At higher temperatures, nucleation is less likely and does not occur within accessible simulation times (hundreds of ns). At lower temperatures, either multiple nucleation events may occur simultaneously without an induction period due to the increased nucleation probability, or diffusion slows down to the extent that nucleation does not occur within accessible simulation times (shown later in Section \ref{sec:path}).
In the regime where a single nucleation event can be clearly identified within simulation times of hundreds of nanoseconds (as in Fig. \ref{fig:potencialenergy}), the nucleation rate is given by \cite{camarillo2024effect}:

\begin{equation}
    J_{het}=\frac{1}{t_{n}A},
    \label{eq:ratesimulation}
\end{equation}
where $A$ is the substrate area in contact with
the liquid and $t_n$
is the time at which nucleation takes place 
averaged over independent trajectories. 
\textcolor{black}{This expression is routinely used to calculate spontaneous nucleation rates \cite{filion2010crystal,espinosa2016seeding,Yuhara2015a} (substituting the
area by the volume if one deals with homogeneous instead of heterogeneous nucleation).}
The nucleation time for an individual 
trajectory is identified as the time at which 
the potential energy begins to drop. 
\textcolor{black}{As shown in Fig. \ref{fig:potencialenergy} for $\delta =6$, 
we typically employ 5 trajectories and wait for nucleation to occur
in all of them to compute the rate via Eq. \ref{eq:ratesimulation}. The 
statistical error in the $\log_{10}J_{het}$
data
reported in Table \ref{tab:rate-results}
comes from assigning to $J$ the same relative error of the nucleation time, which is obtained
through the standard error of $t_n$ in the 5 independent trajectories. 
This statistical error contributes with $\sim$ 0.2 units to $\log_{10}J$.
In addition, there could be a systematic error associated to the criterion used to identify
the appearance of the nucleus. Here we use the potential energy drop previously
discussed. Other criteria to detect nucleation could 
lead to a systematic shift in $t_n$.
For instance, one could detect the number of molecules in the 
largest cluster of ice-like molecules, $N_b$, and identify $t_n$ with the moment in which 
a sharp increase of such a number of molecules takes place. 
Both criteria are compared in the bottom panel of Fig. \ref{fig:potencialenergy}
for a selected nucleation trajectory on a $\delta=5$ substrate. 
Clearly, the nucleation times inferred by either criterion are fully consistent with each other.
Moreover, in our previous work with mW we showed that 
a global order parameter is also fully consistent with the potential
energy criterion \cite{camarillo2024effect}.
Therefore, 
the systematic shift in $J$ due to the choice of the nucleation 
detection criterion is negligible compared to the statistical error 
previously discussed. An error bar of 0.2 units of $\log_{10}J$ is then 
a fair description of the uncertainty of our rate calculations.}

Since the area of the substrate in contact with the liquid 
is of the same order in all
systems (45-50 nm$^2$) and the simulation time is also of the same 
order in all cases ($\sim$ 500 ns), the nucleation rate that can be
probed is fixed by the simulation setup to $\sim$ 10$^{23}$s$^{-1}$m$^{-2}$
\textcolor{black}{according to Eq. \ref{eq:ratesimulation}}.
Hence, what changes from one system to another is not the nucleation
rate (reported in Table \ref{tab:rate-results}), but rather the temperature at which nucleation is observed at 
such rate, which we call the nucleation temperature, T$_n$ (reported
in Table \ref{tab:rate-results}). 
\textcolor{black}{It is important to realize that, because the 
rate depends on temperature, T$_n$ depends on the 
rate we have access to. If the accessible rate were lower, T$_n$ would be higher.
That the nucleation temperature is dependent on the 
working set up is not an issue exclusive to simulations. In experiments, the accessible
rate (and, hence, nucleation temperature) is also fixed by the system's volume. In fact, in experiments of homogeneous
ice nucleation the nucleation temperature is around 235 K when microscopic
drops are used, whereas it goes down to 210 K when 
nanoscopic drops are employed \cite{espinosa2018homogeneous}.
This can be understood because large drops give access to small rates 
(the volume is in the denominator of Eq. \ref{eq:ratesimulation}) and vice versa. 
Small rates correspond to high temperatures, and vice versa. 
In the context of this work, the importance of T$_n$ is enabling the comparison 
of the nucleation
efficiency of different substrates (provided
that such a comparison is established for the same $J$). The higher is T$_n$, the more efficient
is the substrate in nucleating ice. }

\textcolor{black}{In rigor, T$_n$ must be obtained for different
substrates for the exact same nucleation rate, which is an unachievable 
task. 
In our previous work with mW \cite{camarillo2024effect}, we narrowed the
$J$ range 
by computing the rate via Eq. \ref{eq:ratesimulation} 
for different temperatures for each substrate type and interpolating T$_n$ for a selected rate. 
This approach is not viable with TIP4P/Ice because it is much 
slower to simulate than mW (about 2 orders of magnitude slower).
In this work we
perform a single rate calculation 
for each substrate. As explained before, by construction, the
calculated rate will be of the same order of magnitude in all cases. 
In practice, to identify T$_n$ we run simulations of a few hundreds
of ns for increasingly
lower temperatures until we identify a temperature where there is a nucleation 
event such as those shown in Fig. \ref{fig:potencialenergy} (induction period and 
potential energy jump). Then, we run several 
trajectories at the temperature identified as $T_n$ (typically 5)  
to obtain the rate with a higher accuracy.}

\textcolor{black}{In this way we compare T$_n$ of different substrates
for $J$'s within the same order of magnitude,  $J \sim$ 10$^{23}$s$^{-1}$m$^{-2}$. 
Indeed, the rates reported in Table \ref{tab:rate-results}
for different substrates lie within the same order of magnitude. 
This approximation is sufficiently good for our purpose of comparing 
nucleation temperatures, particularly taking 
into account that the rate changes very 
sharply with temperature, 
as shown in Fig. 4 of our previous work with mW \cite{camarillo2024effect}.
Based on the $J(T)$ slope obtained in Ref. \cite{camarillo2024effect}, we estimate that the uncertainty of $T_n$  is $\pm 0.5$ K. Such an error bar compensates for the
fact that the nucleation temperatures for different substrates are not compared at exactly the same 
$J$ (although always within the same order of magnitude).}

The two systems with different liquid depths employed for a $\delta = 5$
substrate (see table \ref{tab:systems_parametres}) 
show nucleation at the same temperature,   
\textcolor{black}{and their respective nucleation times have been combined to obtain
the rate reported in Table \ref{tab:rate-results}.
We have also performed simulations with the same liquid depth but larger substrate areas, obtaining consistent nucleation behavior 
(see Section \ref{sec:path}). These observations indicate that the nucleation process investigated here is robust with respect to the simulation-cell dimensions and provide evidence that possible artifacts associated with periodic images and the PME treatment of the electrostatic interactions do not significantly affect our results.}

\begin{figure}[h] 
\includegraphics[width=0.48\textwidth]{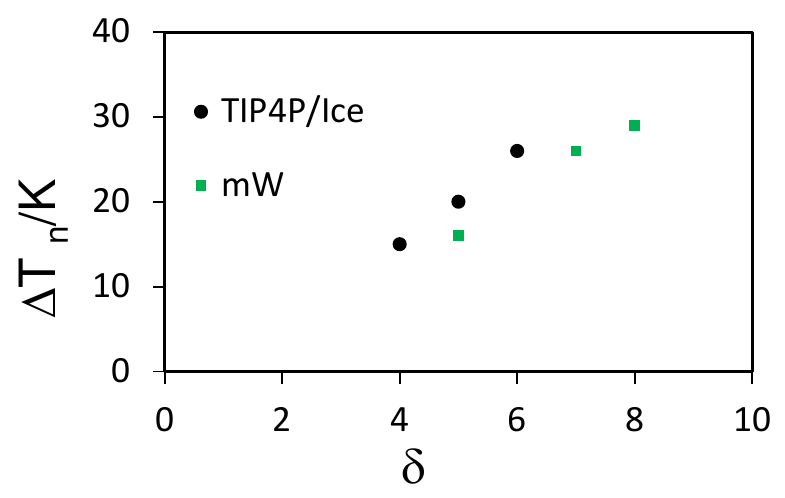} 
    \caption{Subcooling at which nucleation occurs, $\Delta T_n$, versus the stretching percentage, $\delta$, of the ice-like substrate for mW (from Ref. \cite{camarillo2024effect})
    and  TIP4P/Ice (this work). \textcolor{black}{Errors are the size of the symbols.} 
    The pI is exposed to the liquid and the nucleation rate
    is log[J/(m$^{-2}$s$^{-1})]=23\pm$1 in all cases.}
    \label{fig:deltavsdeltaT}
\end{figure}

In Fig. \ref{fig:deltavsdeltaT} we plot $\Delta T_n$, the difference between 
the melting temperature and T$_n$, versus $\delta$
for the TIP4P/Ice model (this work) and compare the data with the results
obtained using the mW model in our previous work \cite{camarillo2024effect}.
In both cases the pI plane of an stretched
ice-like substrate was exposed to the liquid.
\textcolor{black}{
The data corresponding to mW were obtained for a substrate whose molecules
vibrate around their lattice positions. In contrast, in this work molecules of the substrate
have fixed positions and orientations. In any case, for substrates exposing the pII plane we 
obtained very similar T$_n$'s (within 1,5 K) for rigid and for vibrating substrates
\cite{camarillo2024effect}.
Therefore, it is worth comparing mW and TIP4P/Ice in Fig. \ref{fig:deltavsdeltaT}.}
This figure shows that 
the drop in nucleation temperature with the stretching of ice-like
substrates
\textcolor{black}{is similar for both} models.
\textcolor{black}{The similarity of the linear dependencies obtained with both models, together with the consistency in the estimated values of $T_n$, increases our confidence in the reliability of the results despite the limited range of accessible $\delta$ values. As discussed in Section \ref{sec:path}, this range is restricted at low $\delta$ by the formation of a thick network 
of ice clusters that narrows the liquid region of the system, and at high $\delta$ by the slow dynamics of the TIP4P/Ice model. Although the dynamics remain fast in the mW model, we found that, for sufficiently high $\delta$, liquid molecules penetrate the substrate \cite{camarillo2024effect}.}
\textcolor{black}{The slopes for TIP4P/Ice and mW obtained from linear
fits are 5.5$\pm$0.3 K/$\delta$ and  4.4$\pm$0.5 K/$\delta$, respectively. Roughly,} the effect of stretching the lattice by 1 \% \textcolor{black}{is} lower\textcolor{black}{ing} the
nucleation temperature by \textcolor{black}{5-6 K using a rigid TIP4P/Ice
ice-like substrate exposing the pI plane. 
With mW, using a vibrating substrate that exposes the same plane,
a similar but somewhat smoother effect of
the lattice stretching is obtained: 4-5 K/$\delta$. In Ref. \cite{camarillo2024effect}
we showed that this trend is approximately the same regardless the exposed
plane or the ability of the lattice to vibrate. In this work we also show that
the decrease of $T_n$ with $\delta$ is not strongly dependent on the model potential either.}

Although the $\Delta T_n(T)$ trend is
similar in both models, the TIP4P/Ice data \textcolor{black}{seem to} lie above those
corresponding to mW (see Fig. \ref{fig:deltavsdeltaT}).
\textcolor{black}{For $\delta=5$, where we have direct 
calculations for both models, we have a $\Delta T_n=16$ K for mW and a $\Delta T_n=20$ K for TIP4P/Ice. The 4 K difference in supercooling is outside the error bar of T$_n$  calculations.
However, the fact that the mW substrate is flexible and the TIP4P/Ice is rigid could partially 
compensate for
such a difference, given that rigidity increases the supercooling in the mW model
by 1-2 K \cite{camarillo2024effect}.
In any case, because we are working with idealized generic substrates, our goal is not to quantify the differences between the two models with high precision. Instead, it is sufficient for our purposes to conclude that both TIP4P/Ice and mW predict that, for each 1\% increase in lattice stretch, ice-like substrates become approximately 4–6 K less effective at promoting ice nucleation.}

\textcolor{black}{In Ref. \cite{camarillo2024effect} we 
investigated the effect of compressing 
the substrate lattice. Our results suggested that there is an approximate symmetry between expanding and compressing the substrate lattice: a positive value of $\delta$ produces a decrease in $T_n$ similar in magnitude to that caused by the corresponding negative value of $\delta$. Since we also observe a strong similarity between the TIP4P/Ice and mW models, we expect the symmetry between expansion and compression of the substrate lattice to hold for TIP4P/Ice as well, although this remains to be confirmed with TIP4P/Ice simulations.}

\subsection{Free energy barrier and contact angle}

\textcolor{black}{Following our previous work \cite{camarillo2024effect},
we use an approach which we call J-CNT because it combines calculations of J with the Classical Nucleation Theory (CNT) theoretical framework \cite{gibbsCNT1,gibbsCNT2,volmerweber1926,becker-doring}, 
to provide}
estimates of the heterogeneous nucleation free energy
barrier, $\Delta G_{het}^c$.
\textcolor{black}{We start from} the following CNT expression
for $J_{het}$:
\begin{equation}
    J_{het}=\rho_s f^+_{het}Z_{het} \exp[-\Delta G^c_{het}/(k_BT)]
    \label{eq:Jhet}
\end{equation}
Following our previous work \cite{camarillo2024effect}, we define 
$\rho_s$
 as the number density of molecules in the substrate plane in contact with the liquid. Each of these surface molecules can act as a heterogeneous nucleation site (in analogy with bulk liquid molecules acting as homogeneous nucleation sites in the corresponding density factor for homogeneous nucleation, given by the number density of the bulk liquid).
For $f^+_{het}$ and $Z_{het }$, the attachment rate and the Zeldovich factor, respectively, we use, as initial guesses, the corresponding values
for homogeneous nucleation, $f^+_{hom}$ and $Z_{hom}$,  interpolated  from our previous work using the same water model \cite{espinosa2016time,espinosa2014homogeneous}. 
This enables us to obtain a first estimate of $\Delta G^c_{het}$ via Eq. \ref{eq:Jhet}.

 Then, we use the CNT 
extension to heterogeneous nucleation \cite{volmer1929keimbildung,turnbull1950kinetics}
to estimate the contact angle, $\theta$, of the ice spherical cap nucleus on each 
substrate:
\begin{equation}
    \Delta G_{het}^c=\Delta G_{hom}^c f(\theta).
\end{equation}
$\Delta G_{hom}^c$ is known from our previous work 
on homogeneous nucleation with the TIP4P/Ice water model \cite{espinosa2016time}. $f(\theta)$ is the ratio between the volume of the heterogeneous
critical nucleus and that of the homogeneous one (both have the same radius of
curvature) \cite{volmer1929keimbildung,turnbull1950kinetics} and is given 
by the following expression: 
\begin{equation}
f(\theta) = \frac{(2 + \cos\theta)(1 - \cos\theta)^2}{4}.
\end{equation}
Once we know $\theta$ we can use $f^+_{het}=f^+_{hom}(1-\cos(\theta))/2$ and
$Z_{het}=Z_{hom}/\sqrt{f(\theta)}$, as we suggested in Ref. \cite{camarillo2025heterogenous}, to refine the values of $f^+_{het}$ and $Z_{het}$ used in Eq. \ref{eq:Jhet} to obtain $\Delta G^c_{het}$. 
This process is repeated iteratively until convergence. The values
of the parameters obtained through this scheme are reported in Table \ref{tab:rate-results}. The initial values employed for $f^+$ and $Z$, 
$f^+_{hom}$ and $Z_{hom}$ are also reported for convenience. We already
get nearly convergent values for $\Delta G^c_{het}$ in the first iteration 
because the product of the attachment rate and the Zeldovich factor does
not change much from homogeneous to heterogeneous nucleation given that
the former factor decreases from 
homogeneous to heterogeneous case, whereas the latter increases. Hence, the product 
remains within the same order of magnitude. 
Moreover, since the estimation 
of the nucleation free-energy barrier involves the natural logarithm of 
this product, a high degree of accuracy in $f^+_{\mathrm{het}} Z_{\mathrm{het}}$ is not required
to get $\Delta G^c_{het}$ through Eq. \ref{eq:Jhet}.

The values of $\Delta G^c_{het}$ obtained through \textcolor{black}{the J-CNT} 
scheme 
are reported in Table \ref{tab:rate-results}. 
In all TIP4P/Ice cases, barriers of about 11-12 $k_BT$ are obtained, 
which is a reasonable value for a process that takes place
spontaneously during the simulation \cite{camarillo2025heterogenous,filion2010crystal}. 

\begin{figure}
    \centering
    \includegraphics[width=\linewidth]{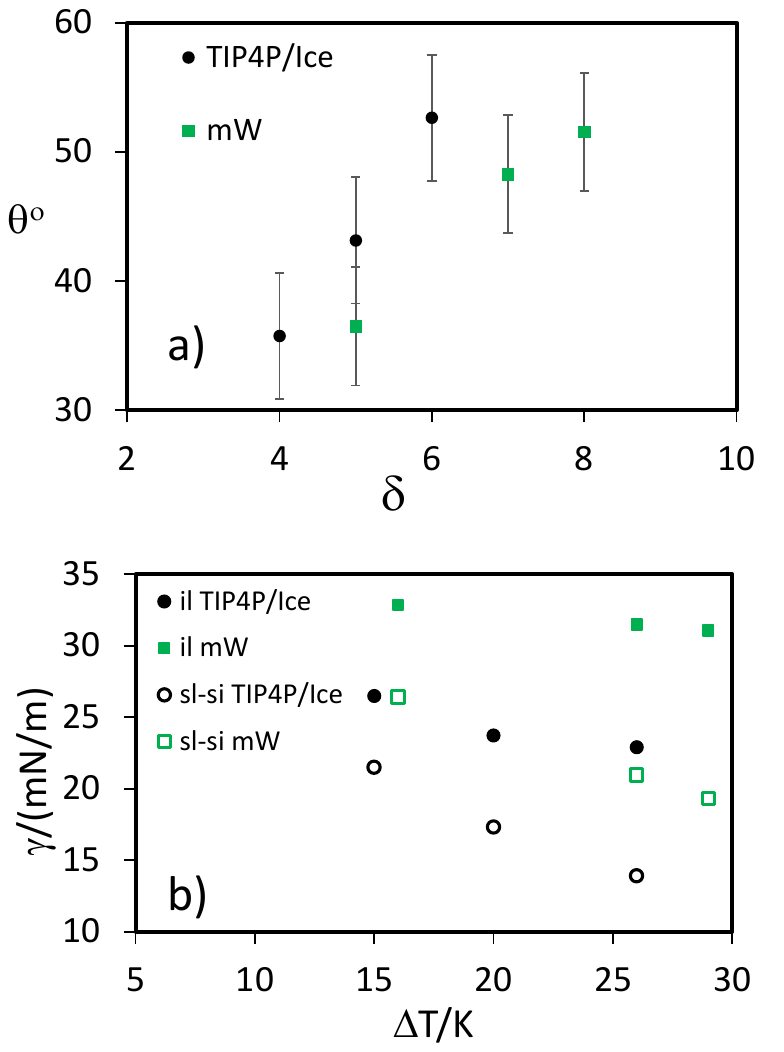}
    \caption{(a) Contact angle in degrees versus the stretching percentage 
    for mW (from Ref. \cite{camarillo2025heterogenous}) and TIP4P/Ice (this work) models (see legend). Substrates
    are stretched ice lattices exposing the pI plane
    to the liquid. (b) Interfacial free energy versus 
    subcooling. Black and green symbols correspond to
    TIP4P/Ice and mW models, respectively. 
    Solid symbols correspond to 
    the ice-liquid interfacial free energy (il), \textcolor{black}{from Ref. \cite{espinosa2016time}}. Empty symbols corresopnd to the difference between the 
    substrate-liquid (sl) and substrate-ice (si) interfacial 
    free energies, \textcolor{black}{obtained in this work through 
    the J-CNT analysis of our rate calculations and Eq. \ref{eq:costheta}}.}
    \label{fig:theta}
\end{figure}

\textcolor{black}{The J-CNT analysis} provides not only $\Delta G^c_{het}$
\textcolor{black}{but also} the contact angle, $\theta$.
In Fig. \ref{fig:theta}(a) we plot $\theta$ versus the stretching
percentage $\delta$. 
The contact angles obtained range from 36 to 53
degrees and, as expected, increase monotonically with $\delta$ (the more stretched the substrate is, the higher the contact angle of the ice nucleus).

A comparison of the two models \textcolor{black}{suggests} that the contact angles are higher for TIP4P/Ice, although the respective error bars overlap for a given $\delta$ value. This is consistent with the fact that a larger subcooling \textcolor{black}{appears to be} required for nucleation to occur in TIP4P/Ice, as shown in Fig.~\ref{fig:deltavsdeltaT}. In other words, the heterogeneous nucleation rate \textcolor{black}{appears to be} higher for mW at a given temperature and $\delta$.
\textcolor{black}{However, after accounting for the effect of lattice rigidity in the mW model, which can increase the required supercooling by 1--2 K \cite{camarillo2024effect}, together with a conservative uncertainty margin, we make the more cautious conclusion that the heterogeneous nucleation rate for TIP4P/Ice is either lower than, or statistically indistinguishable from, that for mW.}
This is somewhat surprising given
that, in the subcooling range here investigated
(15-30 K), homogeneous nucleation rates are several
orders of magnitude higher for TIP4P/Ice than for mW \cite{espinosa2016time}.
The larger value of the ice-liquid 
interfacial free energy, $\gamma_{il}$, for mW justifies
its lower homogeneous ice nucleation rate \cite{espinosa2016time}. 
However, also the 
substrate-liquid  and the substrate-ice interfacial
free energies, $\gamma_{sl}$ and $\gamma_{si}$ 
respectively, play a role in heterogeneous nucleation. 
Specifically, the contact angle is given by
Young's equation \cite{young1805iii}:
\begin{equation}
\cos(\theta)=\frac{\gamma_{sl}-\gamma_{si}}{\gamma_{il}}.
\label{eq:costheta}
\end{equation}
Therefore, even though the denominator in Eq. \ref{eq:costheta} is larger for mW, the numerator can compensate,  
leading to a higher $\cos(\theta)$ (lower
contact angle) for mW. 
In Table. \ref{tab:rate-results} we provide 
$\gamma_{il}$ (from Ref. \cite{espinosa2016time}) and
$\gamma_{sl}-\gamma_{si}$.
\textcolor{black}{The latter
is
obtained from the numerator of Eq. \ref{eq:costheta}, 
as a byproduct of the $J-CNT$ analysis}. The values of these interfacial
free energies are also represented in Fig. \ref{fig:theta}(b).
Solid symbols represent $\gamma_{il}$, which
is indeed larger for mW.
Also the numerator of Eq. \ref{eq:costheta},
given by the empty symbols in Fig. \ref{fig:theta}(b), is larger for mW. In the end, 
the overall ratio is higher 
for mW in Eq. \ref{eq:costheta}, leading
to lower contact angles for mW and to a
faster heterogeneous nucleation rate, in contrast
to the homogeneous nucleation case. 
The molecular origin of the lower $\gamma_{sl}-\gamma_{si}$ difference in TIP4P/Ice is not straightforward and cannot be inferred intuitively. Its elucidation would require rigorous free-energy calculations beyond the scope of this work.

\subsection{Nucleation path}
\label{sec:path}
\begin{figure*}[h] 
    \centering
     \includegraphics[width=0.90\textwidth]{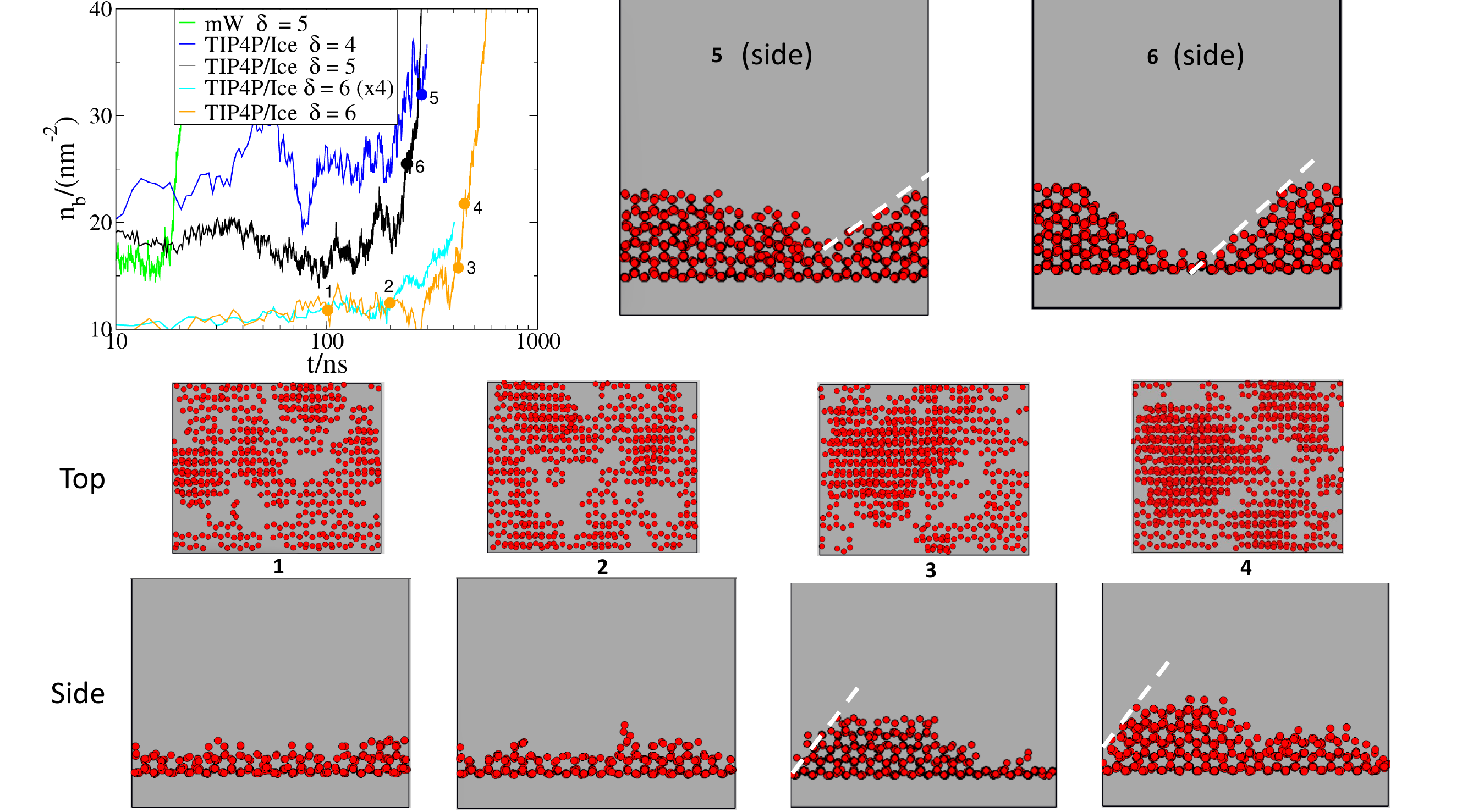}
    \caption{
    \textcolor{black}{Top left: number of molecules in the biggest ice cluster normalized by the substrate area, n$_b$,
    on substrates with different stretching percentages (see legend). 
    For $\delta=5$, we show a comparison with mW
    (green curve), re-analyzed from Ref. \cite{camarillo2024effect}
    For $\delta = 6$, we include a comparison with a system having a four-fold increase in substrate area (x4), constructed by replicating the original system $1\times2\times2$ times (cyan curve).
    The $T_n$ temperature corresponding to each $\delta$
    is reported in table \ref{tab:rate-results}.
    Numbered circles
    correspond to the points where snapshots of the biggest ice cluster are shown.
    Top and/or side projections of the oxygen atoms (red circles) are presented as indicated
    in the figure. The white dashed line in plots where the side view of a nucleus is presented depicts the contact angle reported in Table \ref{tab:rate-results} for the corresponding $\delta$.}}
    \label{fig:nbbigestnormalised}
\end{figure*}

We identify, as described in Section \ref{sec:method}, the 
biggest cluster of ice-like molecules in the liquid slab in order
to have a visual representation of the nucleation path. 
In \textcolor{black}{the top left panel} of Fig.~\ref{fig:nbbigestnormalised} we show the number of molecules in the 
largest ice cluster \textcolor{black}{divided by the substrate area, $n_b$, }as a function of time,
for \textcolor{black}{several} nucleation trajectories.
\textcolor{black}{In all cases,} after a long induction period during which the 
cluster size fluctuates, the number of molecules in the cluster increases
abruptly. This behavior is typical of nucleation followed by growth. 

\textcolor{black}{We analyze in some detail the trajectory for $\delta = 6$, corresponding to the orange curve in Fig.~\ref{fig:nbbigestnormalised}. The numbered symbols along the orange curve indicate the time instants at which snapshots of the largest ice cluster are shown. Only oxygen atoms belonging to the largest ice cluster are represented (as red spheres).
Snapshots corresponding to points 1 and 2 show pre-critical clusters
collected during the induction period, where $n_b$ fluctuates around an average value.
 Top and side views of the clusters are provided in the figure. Whereas the side view gives the impression of a nearly continuous ice layer induced by the substrate, the top view reveals that the largest cluster actually consists of a percolating network of interconnected subcritical ice-like domains separated by regions of lower density. This morphology is consistent with the CNT picture of a fluctuating population of subcritical clusters, one of which eventually reaches the critical size through a rare fluctuation. Therefore, the substrate-induced ice-like region should not be regarded as a homogeneous wetting layer but rather as a heterogeneous network of subcritical ice-like domains.}

\textcolor{black}{To assess possible system-size effects, we performed additional simulations using a $1\times2\times2$ replication of the original $\delta=6$ system. The substrate--liquid contact area, and consequently the number of molecules, is therefore four times larger than in the original system. The analysis of one of the nucleation trajectories obtained for the replicated system is shown in cyan in the top-left panel of Fig. \ref{fig:nbbigestnormalised}. 
Visual inspection of 
the nucleation mechanism in this large system also reveals that pre-critical ice-like order
is due to a network of sub-critical clusters (not shown).
According to the top left panel of Fig. \ref{fig:nbbigestnormalised}, the density of ice-like molecules contained in the cluster network during the induction period, $n_b$, is essentially the same in the two systems, indicating that this quantity is an equilibrium property that is not affected by the system size. 
The nucleation time is also comparable in the two trajectories, with nucleation occurring earlier in the larger system, as expected from the fact that the average nucleation time decreases with increasing substrate area. The consistency found between the original and the four-fold replicated
system gives us confidence in the absence of finite-size effects in our simulations.}

\textcolor{black}{The top left panel of Fig. \ref{fig:nbbigestnormalised} 
also shows that as
the lattice mismatch decreases, $n_b$ increases markedly, indicating that substrates with a better structural match to ice stabilize a larger amount of ice-like order
--denser and thicker networks of subcritical clusters-- during the induction period.
Approximately, for $\delta=4$ $n_b$ is about 2-3 times larger than for the $\delta=6$ substrate
(the mean $n_b$ values during the induction period are 26(1), 17(1) and 12(1) nm$^{-2}$ for $\delta=4, 5$ and 6 respectively).
In our previous work with the mW model \cite{camarillo2024effect}, we also observed that the substrate 
induces sub-critical nuclei during the induction period (instead of a homogeneous 
ice layer).
In fact, according to the similar height of the mW and TIP4P/Ice curves 
during the induction period shown in Fig.~\ref{fig:nbbigestnormalised}
($n_b$ is 17(1) nm$^{-2}$ and 16(1) nm$^{-2}$ for TIP4P/Ice and mW, respectively), the amount of ice-like order induced by the substrate does not strongly depend on the interaction potential.
It does depend much more strongly on the degree of similarity between the substrate and ice.}

\textcolor{black}{
Snapshots 3 and 4 in Fig. \ref{fig:nbbigestnormalised} correspond to post-critical growing clusters taken during the rapid growth regime on the $\delta=6$ substrate, where $n_b$ increases sharply. The dashed lines in the side-view projections indicate the contact angle predicted by our J-CNT analysis. In both snapshots, taken at different stages of growth, a contact angle of $\theta = 53^\circ$ provides a reasonable description of the side profile of the growing cluster.
Snapshots 5 and 6 correspond to representative growing clusters for the $\delta = 4$ and $\delta = 5$ trajectories, respectively.
Again, the white dashed lines indicating the contact angle obtained from the J-CNT analysis provide a reasonable description of the side profiles of the growing clusters, which in all cases can be regarded as thermal fluctuations around an average spherical-cap shape. A more detailed statistical characterization of the critical cluster morphology would certainly be desirable in future work, requiring, for instance, a full committor analysis \cite{dellago2002transition,moroni2005interplay} to identify 
an ensemble of critical nuclei. Nevertheless, the fact that the contact angles obtained from the J-CNT analysis are compatible with the microscopic appearance of the growing clusters lends support to the applicability of CNT to describe the morphology of heterogeneous ice nuclei on the substrates considered here.}

\begin{figure}[h] 
    \centering
    \includegraphics[width=0.48\textwidth]{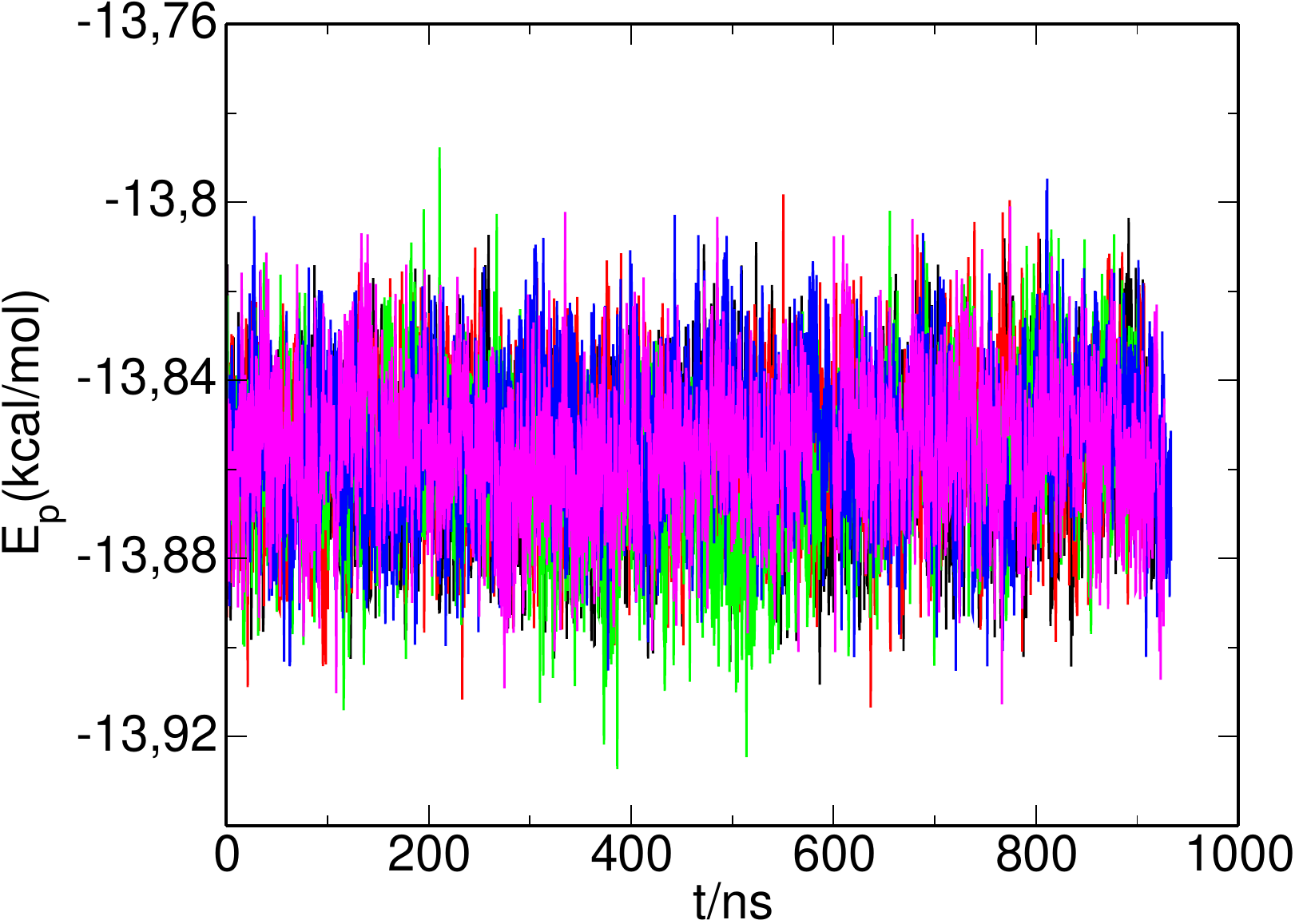} 
    \caption{Potential energy vs time for TIP4P/Ice ice-like substrates stretched by 7 \% for 5 different trajectories at 240 K.}
    \label{fig:7dist}
\end{figure}

Even though TIP4P/Ice and mW show a similar $n_b$ in Fig. \ref{fig:nbbigestnormalised}
and a similar $\Delta T_n$ trend in Fig. \ref{fig:deltavsdeltaT}, 
there are differences in the physics of the mW and TIP4P/Ice models at deep 
subcooling. Whereas we were able to evaluate heterogeneous nucleation rates for 
ice-like substrates with an 8\% stretch using the mW model 
\cite{camarillo2024effect}, we were unable to go beyond
6\%
with 
TIP4P/Ice \textcolor{black}{due to slow dynamics}. 
\textcolor{black}{An extrapolation of a linear fit to the T$_n$ vs $\delta$ data
for TIP4P/Ice predicts that nucleation would occur at a subcooling of
$\sim$31--32 K for $\delta = 7$. At this subcooling, the ratio of the
diffusion coefficients of the mW and TIP4P/Ice models
\cite{espinosa2014homogeneous,espinosa2016time} is $\sim$50.
Therefore, simulations approximately 50 times longer would be required for
TIP4P/Ice to observe the same phenomenon, namely nucleation, making such
simulations computationally too expensive to obtain nucleation events for
TIP4P/Ice at this subcooling.}
This impossibility is illustrated in 
Fig.~\ref{fig:7dist} where we show, for a
$\delta = 7$ substrate at a 
subcooling of 30~K, 
the potential energy as a function of 
time for several trajectories using 
the TIP4P/Ice model. No nucleation events were observed even after simulating 
nearly 1~$\mu$s per trajectory. The significantly larger diffusivity of the mW model compared to that of TIP4P/Ice 
\cite{espinosa2016time} provides a plausible explanation for this different 
behavior, which is 
consistent with the bulk scenario. In bulk simulations, spontaneous
homogeneous 
ice nucleation  has been reported for the mW model \cite{moore2011structural}, but it has never been achieved for TIP4P/Ice.

\textcolor{black}{Therefore, the accessible range of $\delta$ values for the TIP4P/Ice model
is limited at high $\delta$ by slow diffusion and at low $\delta$ by the induction
of a significant degree of ice ordering by the substrate. According to the $n_b$
value shown in the top left panel of Fig.~\ref{fig:nbbigestnormalised}, for
$\delta = 4$ approximately 1200 out of the 5120 water molecules in the system already
exhibit ice-like local order. For this reason, we have restricted our study to
$\delta = 4$, 5, and 6. The present data do not allow us to assess whether deviations from linearity may arise outside the accessible $\delta$ range.}

\begin{table*}[h!]
\centering
\resizebox{\textwidth}{!}{
\begin{tabular}{lcccccccccccccc}
\toprule
Model & $\delta$ & T$_n$/K & $\log_{10}(J_{\mathrm{het}}/(m^{-2}s^{-1}))$ & $\rho_s$/(nm$^{-2}$) & 10$^{-11}$$\cdot$f$^+_{het}$/s$^{-1}$ & 10$^{3}$$\cdot$Z$_{het}$ & $\Delta G^c_{het}$/($k_B$T) & $\Delta G^c_{hom}$/($k_B$T) & $\theta$/$^o$ & 10$^{-11}$$\cdot$f$^+_{hom}$/s$^{-1}$ & 10$^{3}$$\cdot$Z$_{hom}$ & $\Delta$T$_n$/K & $\gamma_{il}$/(mN/m) & $(\gamma_{sl}-\gamma_{si})$/(mN/m)\\
\hline
TIP4P/Ice & 4 & 255 & 23.3 & 5.51 & 5.9 & 5.6 & 11.4 & 457 & 36 & 62 & 0.89 & 15 & 26.5 & 21.5 \\
TIP4P/Ice & 5 & 250 & 23.3  & 5.40 & 3.9 & 8.6 & 11.4 & 229 & 43 & 29 & 1.9 & 20 & 23.7 & 17.3 \\
TIP4P/Ice & 6 & 244 & 22.7 & 5.30 & 2.2 & 10 & 12.3 & 122 & 53 & 11 & 3.2 & 26 & 22.9 & 13.9 \\
mW         & 5 & 257 & 23.6  & 5.67 & 540 & 5.9 & 16.0 & 592 & 36 & 5500 & 0.96 & 16 & 32.8 & 26.4 \\
mW         & 7 & 247 & 23.6  & 5.46 & 260 & 9.9 & 15.7 & 211 & 48 & 1500 & 2.7 & 26 & 31.5 & 20.9 \\
mW         & 8 & 244 & 23.6  & 5.36 & 210 & 11 & 15.7 & 167 & 52 & 1100 & 3.4 & 29 & 31.1 & 19.3 \\
\hline
\hline
\label{tab:rate-results}
\end{tabular}}
\caption{Nucleation parameters for the models considered 
in this work. All TIP4P/Ice heterogeneous nucleation data 
were obtained in this work. The nucleation temperature, T$_n$, and the rate, $J$, of 
the mW model (pI plane) were taken from Ref. \cite{camarillo2024effect}, whereas related heterogeneous nucleation parameters
are reported here for the first time. Homogeneous nucleation
data, reported for completeness, were obtained in previous work \cite{espinosa2016time}. The $\gamma_{il}$ for mW was obtained
from the fit reported in Ref. \cite{espinosa2016time} using 
the more recent value of 273 K \cite{hudait2016free} as the model's melting temperature.}
\label{tab:rate-results}
\end{table*}

\begin{figure}[h] 
    \centering
    \includegraphics[width=0.48\textwidth]{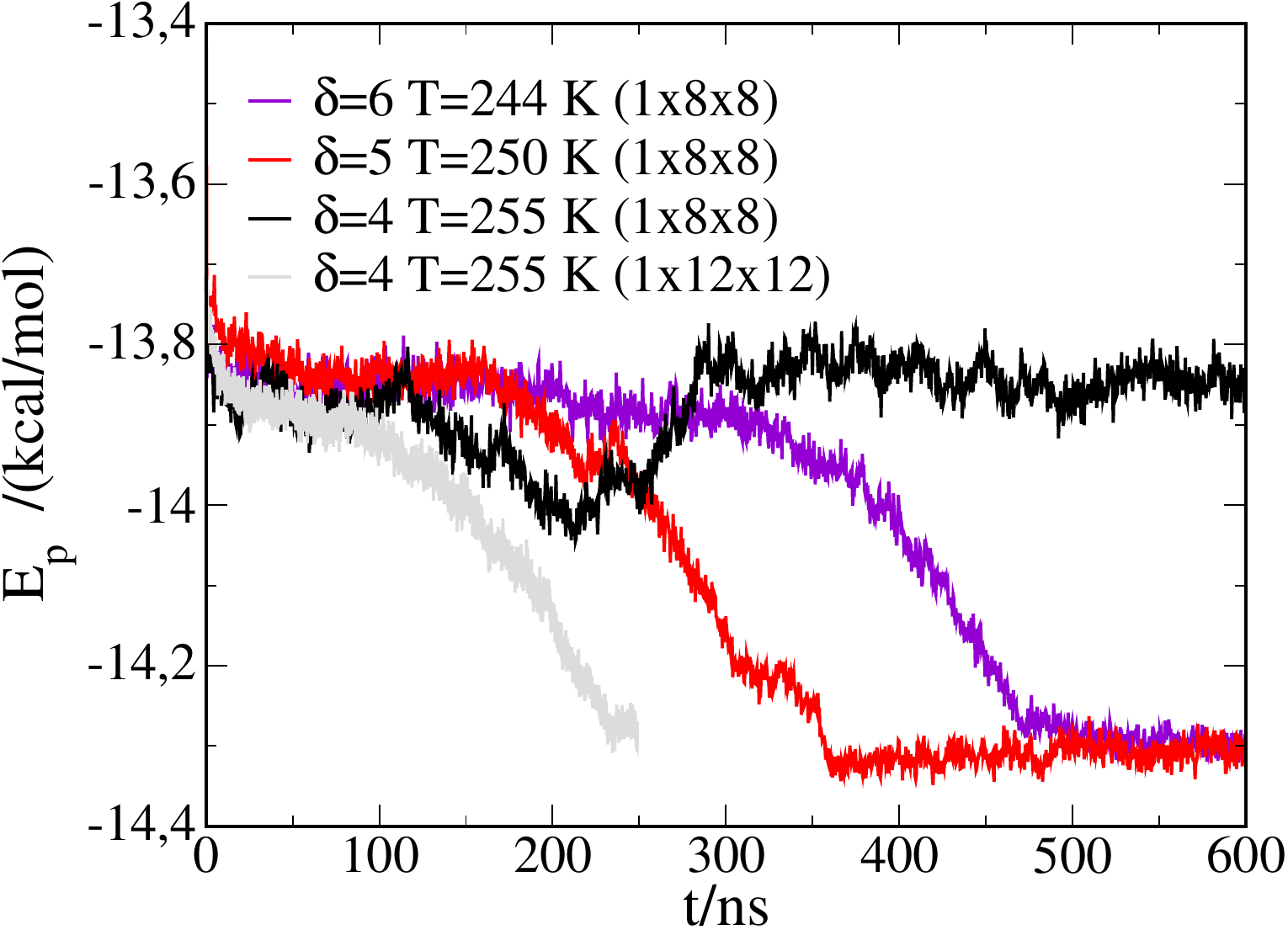} 
    \caption{Potential energy versus time for representative nucleation trajectories
    on substrates with different $\delta$'s and different 
    sizes (as indicated by the unit cell replication factors in the legend). For $\delta$=5 (in red) and $\delta$= 6 (in purple) 
    the potential energy reaches a final plateau (full crystallization) whereas for the small $\delta$=4 system (in black) the ice front melts back after nucleation and initial growth. A larger
    system for $\delta = 4$ (in gray) does show full crystallization.}
    \label{fig:456evstime}
\end{figure}

Finally, we discuss in this section the growth of the ice nucleus. 
The potential energy time evolution for representative nucleation
trajectories in substrates with different $\delta$'s is represented in Fig. \ref{fig:456evstime}. The behaviour corresponding to the trajectories on $\delta$=5 and $\delta=6$
substrates --sudden energy drop that reaches a final plateau-- corresponds to 
nucleation and full growth of the ice phase 
(as shown also in Fig. \ref{fig:potencialenergy} for different trajectories with $\delta =6$). However, the behaviour of trajectories on $\delta=4$ substrates is
qualitatively different. A sudden drop of the potential energy, corresponding to nucleation,
is clearly seen, but the ice phase does not fully grow. Instead, the ice front melts again and potential energy values characteristic of the
induction period are recovered (see black curve
in Fig. \ref{fig:456evstime}). 
This behaviour can be understood by examining a final configuration after ice growth. 
In Fig. \ref{fig:secuenciafotos}(c) we show a snapshot of one of these configurations
for ice nucleation and growth on a $\delta=6$ substrate. A dislocation in the newly
grown ice can be appreciated. Such an extra column of hexagons 
fills the space along the pbc's parallel to the 
interface. Therefore, the dislocation is a consequence of both the unit cell of the
growing ice being smaller than that of the substrate and 
the application of pbc's. 
In the system with the smallest mismatch, $\delta$=4, due to the 
small difference between the substrate and the ice unit cells, there is not enough
space to accommodate an extra column of hexagons 
and the growing ice cannot find a solution to fit across the pbc's. As a consequence, 
the growing ice melts back. 
\textcolor{black}{This finite-size effect affects the subsequent growth of the ice phase, but the
nucleation event itself occurs before the growth incompatibility with the periodic
boundary conditions becomes apparent.}
In fact, we show in Fig. 
 \ref{fig:456evstime} a gray curve corresponding to a $\delta = 4$
 system with a larger substrate area (the unit cell was replicated 1x12x12
 versus 1x8x8 for the smaller system) and crystallization is able to 
 fully proceed as shown by the 
 corresponding energy drop.
\textcolor{black}{The earlier nucleation observed in the $1\times12\times12$ system compared with the
$1\times8\times8$ system is expected from Eq. \ref{eq:ratesimulation}, which implies that
the average nucleation time is inversely proportional to the substrate area.}

\section{Conclusions}
We use molecular simulations with the TIP4P/Ice water model
to investigate the effect of lattice mismatch on heterogeneous 
ice nucleation. We use idealized substrates, with ice Ih structure
and composed of water molecules, to focus on the 
effect of lattice mismatch, which we quantify by 
stretching the ice lattice by a certain factor. 
This study follows our recent work \cite{camarillo2024effect}, where
we used the coarse-grained mW water model.

We place subcooled water in contact with the stretched ice substrate and 
wait for spontaneous ice nucleation events to occur. We search for the 
temperature at which nucleation takes place after an induction 
period of tens-hundreds of nanoseconds. We define this temperature 
as the nucleation temperature,  $T_n$, and we observe that it roughly decreases by \textcolor{black}{5-6 K} per percentage point of ice lattice stretching, similar \textcolor{black}{but a bit 
larger than the 4-5 K} we found using the
mW model \cite{camarillo2024effect}. 

\textcolor{black}{Beyond showing that the atomistic and coarse-grained models predict a comparable lattice-mismatch effect, this work reveals new physical insights compared with our previous study using the mW model \cite{camarillo2024effect}. We discuss these findings in the following paragraphs.}

In both models, mW and TIP4P/Ice, ice-like substrates induce
\textcolor{black}{a  fluctuating network of ice-like subcritical clusters} in the adjacent liquid
during the induction period, \textcolor{black}{rather than a uniform wetting layer}. 
\textcolor{black}{The density of this network increases as the lattice mismatch decreases and is remarkably similar for the two models at a given value of $\delta$. The critical nucleus subsequently emerges stochastically from this network with a spherical cap shape.} 

Following the approach of Ref. \cite{camarillo2025heterogenous}, we 
combine our heterogeneous nucleation rate calculations with Classical Nucleation Theory
and with previous homogeneous nucleation rate calculations 
to estimate nucleation barriers, kinetic pre-factors and contact angles. 
Interestingly, the contact angle obtained through this macroscopic approach is 
consistent with a microscopic analysis of the nucleation trajectories. 

Heterogeneous ice nucleation \textcolor{black}{appears to be slower for TIP4P/Ice (or, at most, no faster than mW when the effects of substrate rigidity and the uncertainties in the data are taken into account) in the studied subcooling range (15–30 K).}
In  
the case of homogeneous nucleation, it is faster for TIP4P/Ice 
due to its lower ice-liquid interfacial free energy
\cite{espinosa2016time}.
A detailed and rigorous study of the energetic balance of the interfacial
free energies involved in the heterogeneous nucleation process is required
to better understand this comparison.

\textcolor{black}{It would be interesting to extend the present study to next-generation water models, such as AI/ML-based potentials. The close agreement between the coarse-grained mW and atomistic TIP4P/Ice models suggests that the dependence of heterogeneous ice nucleation on substrate lattice strain does not strongly depend on the specific interaction potential. Nevertheless, AI/ML-based models would provide a valuable opportunity to further test the generality of the main conclusions of this study.}


\begin{acknowledgments}
This work was funded by Grants No. 
PID2022-136919NB-C31 and PID2022-136919NB-C32 of the MICINN. The authors gratefully acknowledge the Universidad Politecnica de Madrid (www.upm.es) for providing computing resources on Magerit Supercomputer.  
\end{acknowledgments}

\clearpage
\section{REFERENCES}


\begin{thebibliography}{47}%
\makeatletter
\providecommand \@ifxundefined [1]{%
 \@ifx{#1\undefined}
}%
\providecommand \@ifnum [1]{%
 \ifnum #1\expandafter \@firstoftwo
 \else \expandafter \@secondoftwo
 \fi
}%
\providecommand \@ifx [1]{%
 \ifx #1\expandafter \@firstoftwo
 \else \expandafter \@secondoftwo
 \fi
}%
\providecommand \natexlab [1]{#1}%
\providecommand \enquote  [1]{``#1''}%
\providecommand \bibnamefont  [1]{#1}%
\providecommand \bibfnamefont [1]{#1}%
\providecommand \citenamefont [1]{#1}%
\providecommand \href@noop [0]{\@secondoftwo}%
\providecommand \href [0]{\begingroup \@sanitize@url \@href}%
\providecommand \@href[1]{\@@startlink{#1}\@@href}%
\providecommand \@@href[1]{\endgroup#1\@@endlink}%
\providecommand \@sanitize@url [0]{\catcode `\\12\catcode `\$12\catcode `\&12\catcode `\#12\catcode `\^12\catcode `\_12\catcode `\%12\relax}%
\providecommand \@@startlink[1]{}%
\providecommand \@@endlink[0]{}%
\providecommand \url  [0]{\begingroup\@sanitize@url \@url }%
\providecommand \@url [1]{\endgroup\@href {#1}{\urlprefix }}%
\providecommand \urlprefix  [0]{URL }%
\providecommand \Eprint [0]{\href }%
\providecommand \doibase [0]{http://dx.doi.org/}%
\providecommand \selectlanguage [0]{\@gobble}%
\providecommand \bibinfo  [0]{\@secondoftwo}%
\providecommand \bibfield  [0]{\@secondoftwo}%
\providecommand \translation [1]{[#1]}%
\providecommand \BibitemOpen [0]{}%
\providecommand \bibitemStop [0]{}%
\providecommand \bibitemNoStop [0]{.\EOS\space}%
\providecommand \EOS [0]{\spacefactor3000\relax}%
\providecommand \BibitemShut  [1]{\csname bibitem#1\endcsname}%
\let\auto@bib@innerbib\@empty
\bibitem [{\citenamefont {Pruppacher}, \citenamefont {Klett},\ and\ \citenamefont {Wang}(1998)}]{pruppacher1998microphysics}%
  \BibitemOpen
  \bibfield  {author} {\bibinfo {author} {\bibfnamefont {H.~R.}\ \bibnamefont {Pruppacher}}, \bibinfo {author} {\bibfnamefont {J.~D.}\ \bibnamefont {Klett}}, \ and\ \bibinfo {author} {\bibfnamefont {P.~K.}\ \bibnamefont {Wang}},\ }\href@noop {} {\enquote {\bibinfo {title} {Microphysics of clouds and precipitation},}\ } (\bibinfo {year} {1998})\BibitemShut {NoStop}%
\bibitem [{\citenamefont {Cantrell}\ and\ \citenamefont {Heymsfield}(2005)}]{cantrell2005production}%
  \BibitemOpen
  \bibfield  {author} {\bibinfo {author} {\bibfnamefont {W.}~\bibnamefont {Cantrell}}\ and\ \bibinfo {author} {\bibfnamefont {A.}~\bibnamefont {Heymsfield}},\ }\bibfield  {title} {\enquote {\bibinfo {title} {Production of ice in tropospheric clouds: A review},}\ }\href@noop {} {\bibfield  {journal} {\bibinfo  {journal} {Bulletin of the American Meteorological Society}\ }\textbf {\bibinfo {volume} {86}},\ \bibinfo {pages} {795--808} (\bibinfo {year} {2005})}\BibitemShut {NoStop}%
\bibitem [{\citenamefont {Hoose}\ and\ \citenamefont {M{\"o}hler}(2012)}]{hoose2012heterogeneous}%
  \BibitemOpen
  \bibfield  {author} {\bibinfo {author} {\bibfnamefont {C.}~\bibnamefont {Hoose}}\ and\ \bibinfo {author} {\bibfnamefont {O.}~\bibnamefont {M{\"o}hler}},\ }\bibfield  {title} {\enquote {\bibinfo {title} {Heterogeneous ice nucleation on atmospheric aerosols: a review of results from laboratory experiments},}\ }\href@noop {} {\bibfield  {journal} {\bibinfo  {journal} {Atmospheric Chemistry and Physics}\ }\textbf {\bibinfo {volume} {12}},\ \bibinfo {pages} {9817--9854} (\bibinfo {year} {2012})}\BibitemShut {NoStop}%
\bibitem [{\citenamefont {Sosso}\ \emph {et~al.}(2016)\citenamefont {Sosso}, \citenamefont {Chen}, \citenamefont {Cox}, \citenamefont {Fitzner}, \citenamefont {Pedevilla}, \citenamefont {Zen},\ and\ \citenamefont {Michaelides}}]{sosso2016crystal}%
  \BibitemOpen
  \bibfield  {author} {\bibinfo {author} {\bibfnamefont {G.~C.}\ \bibnamefont {Sosso}}, \bibinfo {author} {\bibfnamefont {J.}~\bibnamefont {Chen}}, \bibinfo {author} {\bibfnamefont {S.~J.}\ \bibnamefont {Cox}}, \bibinfo {author} {\bibfnamefont {M.}~\bibnamefont {Fitzner}}, \bibinfo {author} {\bibfnamefont {P.}~\bibnamefont {Pedevilla}}, \bibinfo {author} {\bibfnamefont {A.}~\bibnamefont {Zen}}, \ and\ \bibinfo {author} {\bibfnamefont {A.}~\bibnamefont {Michaelides}},\ }\bibfield  {title} {\enquote {\bibinfo {title} {Crystal nucleation in liquids: Open questions and future challenges in molecular dynamics simulations},}\ }\href@noop {} {\bibfield  {journal} {\bibinfo  {journal} {Chemical reviews}\ }\textbf {\bibinfo {volume} {116}},\ \bibinfo {pages} {7078--7116} (\bibinfo {year} {2016})}\BibitemShut {NoStop}%
\bibitem [{\citenamefont {Lupi}, \citenamefont {Hudait},\ and\ \citenamefont {Molinero}(2014)}]{lupi2014heterogeneous}%
  \BibitemOpen
  \bibfield  {author} {\bibinfo {author} {\bibfnamefont {L.}~\bibnamefont {Lupi}}, \bibinfo {author} {\bibfnamefont {A.}~\bibnamefont {Hudait}}, \ and\ \bibinfo {author} {\bibfnamefont {V.}~\bibnamefont {Molinero}},\ }\bibfield  {title} {\enquote {\bibinfo {title} {Heterogeneous nucleation of ice on carbon surfaces},}\ }\href@noop {} {\bibfield  {journal} {\bibinfo  {journal} {Journal of the American Chemical Society}\ }\textbf {\bibinfo {volume} {136}},\ \bibinfo {pages} {3156--3164} (\bibinfo {year} {2014})}\BibitemShut {NoStop}%
\bibitem [{\citenamefont {Zielke}, \citenamefont {Bertram},\ and\ \citenamefont {Patey}(2015)}]{zielke2015molecular}%
  \BibitemOpen
  \bibfield  {author} {\bibinfo {author} {\bibfnamefont {S.~A.}\ \bibnamefont {Zielke}}, \bibinfo {author} {\bibfnamefont {A.~K.}\ \bibnamefont {Bertram}}, \ and\ \bibinfo {author} {\bibfnamefont {G.~N.}\ \bibnamefont {Patey}},\ }\bibfield  {title} {\enquote {\bibinfo {title} {A molecular mechanism of ice nucleation on model agi surfaces},}\ }\href@noop {} {\bibfield  {journal} {\bibinfo  {journal} {The Journal of Physical Chemistry B}\ }\textbf {\bibinfo {volume} {119}},\ \bibinfo {pages} {9049--9055} (\bibinfo {year} {2015})}\BibitemShut {NoStop}%
\bibitem [{\citenamefont {Pedevilla}\ \emph {et~al.}(2016)\citenamefont {Pedevilla}, \citenamefont {Cox}, \citenamefont {Slater},\ and\ \citenamefont {Michaelides}}]{pedevilla2016can}%
  \BibitemOpen
  \bibfield  {author} {\bibinfo {author} {\bibfnamefont {P.}~\bibnamefont {Pedevilla}}, \bibinfo {author} {\bibfnamefont {S.~J.}\ \bibnamefont {Cox}}, \bibinfo {author} {\bibfnamefont {B.}~\bibnamefont {Slater}}, \ and\ \bibinfo {author} {\bibfnamefont {A.}~\bibnamefont {Michaelides}},\ }\bibfield  {title} {\enquote {\bibinfo {title} {Can ice-like structures form on non-ice-like substrates? the example of the k-feldspar microcline},}\ }\href@noop {} {\bibfield  {journal} {\bibinfo  {journal} {The Journal of Physical Chemistry C}\ }\textbf {\bibinfo {volume} {120}},\ \bibinfo {pages} {6704--6713} (\bibinfo {year} {2016})}\BibitemShut {NoStop}%
\bibitem [{\citenamefont {Fitzner}\ \emph {et~al.}(2015)\citenamefont {Fitzner}, \citenamefont {Sosso}, \citenamefont {Cox},\ and\ \citenamefont {Michaelides}}]{fitzner2015many}%
  \BibitemOpen
  \bibfield  {author} {\bibinfo {author} {\bibfnamefont {M.}~\bibnamefont {Fitzner}}, \bibinfo {author} {\bibfnamefont {G.~C.}\ \bibnamefont {Sosso}}, \bibinfo {author} {\bibfnamefont {S.~J.}\ \bibnamefont {Cox}}, \ and\ \bibinfo {author} {\bibfnamefont {A.}~\bibnamefont {Michaelides}},\ }\bibfield  {title} {\enquote {\bibinfo {title} {The many faces of heterogeneous ice nucleation: Interplay between surface morphology and hydrophobicity},}\ }\href@noop {} {\bibfield  {journal} {\bibinfo  {journal} {Journal of the American Chemical Society}\ }\textbf {\bibinfo {volume} {137}},\ \bibinfo {pages} {13658--13669} (\bibinfo {year} {2015})}\BibitemShut {NoStop}%
\bibitem [{\citenamefont {Glatz}\ and\ \citenamefont {Sarupria}(2018)}]{glatz2018heterogeneous}%
  \BibitemOpen
  \bibfield  {author} {\bibinfo {author} {\bibfnamefont {B.}~\bibnamefont {Glatz}}\ and\ \bibinfo {author} {\bibfnamefont {S.}~\bibnamefont {Sarupria}},\ }\bibfield  {title} {\enquote {\bibinfo {title} {Heterogeneous ice nucleation: Interplay of surface properties and their impact on water orientations},}\ }\href@noop {} {\bibfield  {journal} {\bibinfo  {journal} {Langmuir}\ }\textbf {\bibinfo {volume} {34}},\ \bibinfo {pages} {1190--1198} (\bibinfo {year} {2018})}\BibitemShut {NoStop}%
\bibitem [{\citenamefont {Cox}\ \emph {et~al.}(2012)\citenamefont {Cox}, \citenamefont {Kathmann}, \citenamefont {Purton}, \citenamefont {Gillan},\ and\ \citenamefont {Michaelides}}]{cox2012non}%
  \BibitemOpen
  \bibfield  {author} {\bibinfo {author} {\bibfnamefont {S.~J.}\ \bibnamefont {Cox}}, \bibinfo {author} {\bibfnamefont {S.~M.}\ \bibnamefont {Kathmann}}, \bibinfo {author} {\bibfnamefont {J.~A.}\ \bibnamefont {Purton}}, \bibinfo {author} {\bibfnamefont {M.~J.}\ \bibnamefont {Gillan}}, \ and\ \bibinfo {author} {\bibfnamefont {A.}~\bibnamefont {Michaelides}},\ }\bibfield  {title} {\enquote {\bibinfo {title} {Non-hexagonal ice at hexagonal surfaces: The role of lattice mismatch},}\ }\href@noop {} {\bibfield  {journal} {\bibinfo  {journal} {Physical Chemistry Chemical Physics}\ }\textbf {\bibinfo {volume} {14}},\ \bibinfo {pages} {7944--7949} (\bibinfo {year} {2012})}\BibitemShut {NoStop}%
\bibitem [{\citenamefont {Turnbull}(1950)}]{turnbull1950kinetics}%
  \BibitemOpen
  \bibfield  {author} {\bibinfo {author} {\bibfnamefont {D.}~\bibnamefont {Turnbull}},\ }\bibfield  {title} {\enquote {\bibinfo {title} {Kinetics of heterogeneous nucleation},}\ }\href@noop {} {\bibfield  {journal} {\bibinfo  {journal} {Journal of Chemical Physics}\ }\textbf {\bibinfo {volume} {18}},\ \bibinfo {pages} {198--203} (\bibinfo {year} {1950})}\BibitemShut {NoStop}%
\bibitem [{\citenamefont {Mithen}\ and\ \citenamefont {Sear}(2014)}]{mithen2014epitaxial}%
  \BibitemOpen
  \bibfield  {author} {\bibinfo {author} {\bibfnamefont {J.}~\bibnamefont {Mithen}}\ and\ \bibinfo {author} {\bibfnamefont {R.}~\bibnamefont {Sear}},\ }\bibfield  {title} {\enquote {\bibinfo {title} {Epitaxial nucleation of a crystal on a crystalline surface},}\ }\href@noop {} {\bibfield  {journal} {\bibinfo  {journal} {Europhysics Letters}\ }\textbf {\bibinfo {volume} {105}},\ \bibinfo {pages} {18004} (\bibinfo {year} {2014})}\BibitemShut {NoStop}%
\bibitem [{\citenamefont {Soni}\ and\ \citenamefont {Patey}(2022)}]{soni2022ice}%
  \BibitemOpen
  \bibfield  {author} {\bibinfo {author} {\bibfnamefont {A.}~\bibnamefont {Soni}}\ and\ \bibinfo {author} {\bibfnamefont {G.}~\bibnamefont {Patey}},\ }\bibfield  {title} {\enquote {\bibinfo {title} {Ice nucleation by the primary prism face of silver iodide},}\ }\href@noop {} {\bibfield  {journal} {\bibinfo  {journal} {The Journal of Physical Chemistry C}\ }\textbf {\bibinfo {volume} {126}},\ \bibinfo {pages} {6716--6723} (\bibinfo {year} {2022})}\BibitemShut {NoStop}%
\bibitem [{\citenamefont {Reinhardt}\ and\ \citenamefont {Doye}(2014)}]{reinhardt2014effects}%
  \BibitemOpen
  \bibfield  {author} {\bibinfo {author} {\bibfnamefont {A.}~\bibnamefont {Reinhardt}}\ and\ \bibinfo {author} {\bibfnamefont {J.~P.}\ \bibnamefont {Doye}},\ }\bibfield  {title} {\enquote {\bibinfo {title} {Effects of surface interactions on heterogeneous ice nucleation for a monatomic water model},}\ }\href@noop {} {\bibfield  {journal} {\bibinfo  {journal} {The Journal of chemical physics}\ }\textbf {\bibinfo {volume} {141}},\ \bibinfo {pages} {084501} (\bibinfo {year} {2014})}\BibitemShut {NoStop}%
\bibitem [{\citenamefont {Fraux}\ and\ \citenamefont {Doye}(2014)}]{fraux2014note}%
  \BibitemOpen
  \bibfield  {author} {\bibinfo {author} {\bibfnamefont {G.}~\bibnamefont {Fraux}}\ and\ \bibinfo {author} {\bibfnamefont {J.~P.}\ \bibnamefont {Doye}},\ }\bibfield  {title} {\enquote {\bibinfo {title} {Note: Heterogeneous ice nucleation on silver-iodide-like surfaces},}\ }\href@noop {} {\bibfield  {journal} {\bibinfo  {journal} {The Journal of chemical physics}\ }\textbf {\bibinfo {volume} {141}},\ \bibinfo {pages} {216101} (\bibinfo {year} {2014})}\BibitemShut {NoStop}%
\bibitem [{\citenamefont {Lupi}\ and\ \citenamefont {Molinero}(2014)}]{lupi2014does}%
  \BibitemOpen
  \bibfield  {author} {\bibinfo {author} {\bibfnamefont {L.}~\bibnamefont {Lupi}}\ and\ \bibinfo {author} {\bibfnamefont {V.}~\bibnamefont {Molinero}},\ }\bibfield  {title} {\enquote {\bibinfo {title} {Does hydrophilicity of carbon particles improve their ice nucleation ability?}}\ }\href@noop {} {\bibfield  {journal} {\bibinfo  {journal} {The Journal of Physical Chemistry A}\ }\textbf {\bibinfo {volume} {118}},\ \bibinfo {pages} {7330--7337} (\bibinfo {year} {2014})}\BibitemShut {NoStop}%
\bibitem [{\citenamefont {Camarillo}\ \emph {et~al.}(2024)\citenamefont {Camarillo}, \citenamefont {Oller-Iscar}, \citenamefont {M~Conde}, \citenamefont {Ram{\'\i}rez},\ and\ \citenamefont {Sanz}}]{camarillo2024effect}%
  \BibitemOpen
  \bibfield  {author} {\bibinfo {author} {\bibfnamefont {M.}~\bibnamefont {Camarillo}}, \bibinfo {author} {\bibfnamefont {J.}~\bibnamefont {Oller-Iscar}}, \bibinfo {author} {\bibfnamefont {M.}~\bibnamefont {M~Conde}}, \bibinfo {author} {\bibfnamefont {J.}~\bibnamefont {Ram{\'\i}rez}}, \ and\ \bibinfo {author} {\bibfnamefont {E.}~\bibnamefont {Sanz}},\ }\bibfield  {title} {\enquote {\bibinfo {title} {Effect of substrate mismatch, orientation, and flexibility on heterogeneous ice nucleation},}\ }\href@noop {} {\bibfield  {journal} {\bibinfo  {journal} {The Journal of Chemical Physics}\ }\textbf {\bibinfo {volume} {160}},\ \bibinfo {pages} {134505} (\bibinfo {year} {2024})}\BibitemShut {NoStop}%
\bibitem [{\citenamefont {Molinero}\ and\ \citenamefont {Moore}(2009)}]{molinero2009water}%
  \BibitemOpen
  \bibfield  {author} {\bibinfo {author} {\bibfnamefont {V.}~\bibnamefont {Molinero}}\ and\ \bibinfo {author} {\bibfnamefont {E.~B.}\ \bibnamefont {Moore}},\ }\bibfield  {title} {\enquote {\bibinfo {title} {Water modeled as an intermediate element between carbon and silicon},}\ }\href@noop {} {\bibfield  {journal} {\bibinfo  {journal} {The Journal of Physical Chemistry B}\ }\textbf {\bibinfo {volume} {113}},\ \bibinfo {pages} {4008--4016} (\bibinfo {year} {2009})}\BibitemShut {NoStop}%
\bibitem [{\citenamefont {Abascal}\ \emph {et~al.}(2005)\citenamefont {Abascal}, \citenamefont {Sanz}, \citenamefont {Garc{\'\i}a~Fern{\'a}ndez},\ and\ \citenamefont {Vega}}]{abascal2005potential}%
  \BibitemOpen
  \bibfield  {author} {\bibinfo {author} {\bibfnamefont {J.}~\bibnamefont {Abascal}}, \bibinfo {author} {\bibfnamefont {E.}~\bibnamefont {Sanz}}, \bibinfo {author} {\bibfnamefont {R.}~\bibnamefont {Garc{\'\i}a~Fern{\'a}ndez}}, \ and\ \bibinfo {author} {\bibfnamefont {C.}~\bibnamefont {Vega}},\ }\bibfield  {title} {\enquote {\bibinfo {title} {A potential model for the study of ices and amorphous water: Tip4p/ice},}\ }\href@noop {} {\bibfield  {journal} {\bibinfo  {journal} {The Journal of chemical physics}\ }\textbf {\bibinfo {volume} {122}},\ \bibinfo {pages} {234511} (\bibinfo {year} {2005})}\BibitemShut {NoStop}%
\bibitem [{\citenamefont {Hess}\ \emph {et~al.}(2008)\citenamefont {Hess}, \citenamefont {Kutzner}, \citenamefont {Van Der~Spoel},\ and\ \citenamefont {Lindahl}}]{hess2008gromacs}%
  \BibitemOpen
  \bibfield  {author} {\bibinfo {author} {\bibfnamefont {B.}~\bibnamefont {Hess}}, \bibinfo {author} {\bibfnamefont {C.}~\bibnamefont {Kutzner}}, \bibinfo {author} {\bibfnamefont {D.}~\bibnamefont {Van Der~Spoel}}, \ and\ \bibinfo {author} {\bibfnamefont {E.}~\bibnamefont {Lindahl}},\ }\bibfield  {title} {\enquote {\bibinfo {title} {Gromacs 4: algorithms for highly efficient, load-balanced, and scalable molecular simulation},}\ }\href@noop {} {\bibfield  {journal} {\bibinfo  {journal} {Journal of chemical theory and computation}\ }\textbf {\bibinfo {volume} {4}},\ \bibinfo {pages} {435--447} (\bibinfo {year} {2008})}\BibitemShut {NoStop}%
\bibitem [{\citenamefont {Spreiter}\ and\ \citenamefont {Walter}(1999)}]{spreiter1999classical}%
  \BibitemOpen
  \bibfield  {author} {\bibinfo {author} {\bibfnamefont {Q.}~\bibnamefont {Spreiter}}\ and\ \bibinfo {author} {\bibfnamefont {M.}~\bibnamefont {Walter}},\ }\bibfield  {title} {\enquote {\bibinfo {title} {Classical molecular dynamics simulation with the velocity verlet algorithm at strong external magnetic fields},}\ }\href@noop {} {\bibfield  {journal} {\bibinfo  {journal} {Journal of Computational Physics}\ }\textbf {\bibinfo {volume} {152}},\ \bibinfo {pages} {102--119} (\bibinfo {year} {1999})}\BibitemShut {NoStop}%
\bibitem [{\citenamefont {Martyna}, \citenamefont {Klein},\ and\ \citenamefont {Tuckerman}(1992)}]{martyna1992nose}%
  \BibitemOpen
  \bibfield  {author} {\bibinfo {author} {\bibfnamefont {G.~J.}\ \bibnamefont {Martyna}}, \bibinfo {author} {\bibfnamefont {M.~L.}\ \bibnamefont {Klein}}, \ and\ \bibinfo {author} {\bibfnamefont {M.}~\bibnamefont {Tuckerman}},\ }\bibfield  {title} {\enquote {\bibinfo {title} {Nos{\'e}--hoover chains: The canonical ensemble via continuous dynamics},}\ }\href@noop {} {\bibfield  {journal} {\bibinfo  {journal} {The Journal of chemical physics}\ }\textbf {\bibinfo {volume} {97}},\ \bibinfo {pages} {2635--2643} (\bibinfo {year} {1992})}\BibitemShut {NoStop}%
\bibitem [{\citenamefont {Montero~de Hijes}\ \emph {et~al.}(2019)\citenamefont {Montero~de Hijes}, \citenamefont {Espinosa}, \citenamefont {Vega},\ and\ \citenamefont {Sanz}}]{montero2019ice}%
  \BibitemOpen
  \bibfield  {author} {\bibinfo {author} {\bibfnamefont {P.}~\bibnamefont {Montero~de Hijes}}, \bibinfo {author} {\bibfnamefont {J.}~\bibnamefont {Espinosa}}, \bibinfo {author} {\bibfnamefont {C.}~\bibnamefont {Vega}}, \ and\ \bibinfo {author} {\bibfnamefont {E.}~\bibnamefont {Sanz}},\ }\bibfield  {title} {\enquote {\bibinfo {title} {Ice growth rate: Temperature dependence and effect of heat dissipation},}\ }\href@noop {} {\bibfield  {journal} {\bibinfo  {journal} {The Journal of Chemical Physics}\ }\textbf {\bibinfo {volume} {151}},\ \bibinfo {pages} {044509} (\bibinfo {year} {2019})}\BibitemShut {NoStop}%
\bibitem [{\citenamefont {Wheeler}\ and\ \citenamefont {Newman}(2002)}]{wheeler2002less}%
  \BibitemOpen
  \bibfield  {author} {\bibinfo {author} {\bibfnamefont {D.~R.}\ \bibnamefont {Wheeler}}\ and\ \bibinfo {author} {\bibfnamefont {J.}~\bibnamefont {Newman}},\ }\bibfield  {title} {\enquote {\bibinfo {title} {A less expensive ewald lattice sum},}\ }\href@noop {} {\bibfield  {journal} {\bibinfo  {journal} {Chemical physics letters}\ }\textbf {\bibinfo {volume} {366}},\ \bibinfo {pages} {537--543} (\bibinfo {year} {2002})}\BibitemShut {NoStop}%
\bibitem [{\citenamefont {Hess}\ \emph {et~al.}(1997)\citenamefont {Hess}, \citenamefont {Bekker}, \citenamefont {Berendsen},\ and\ \citenamefont {Fraaije}}]{hess1997lincs}%
  \BibitemOpen
  \bibfield  {author} {\bibinfo {author} {\bibfnamefont {B.}~\bibnamefont {Hess}}, \bibinfo {author} {\bibfnamefont {H.}~\bibnamefont {Bekker}}, \bibinfo {author} {\bibfnamefont {H.~J.}\ \bibnamefont {Berendsen}}, \ and\ \bibinfo {author} {\bibfnamefont {J.~G.}\ \bibnamefont {Fraaije}},\ }\bibfield  {title} {\enquote {\bibinfo {title} {Lincs: A linear constraint solver for molecular simulations},}\ }\href@noop {} {\bibfield  {journal} {\bibinfo  {journal} {Journal of computational chemistry}\ }\textbf {\bibinfo {volume} {18}},\ \bibinfo {pages} {1463--1472} (\bibinfo {year} {1997})}\BibitemShut {NoStop}%
\bibitem [{\citenamefont {Hess}(2008)}]{hess2008p}%
  \BibitemOpen
  \bibfield  {author} {\bibinfo {author} {\bibfnamefont {B.}~\bibnamefont {Hess}},\ }\bibfield  {title} {\enquote {\bibinfo {title} {P-lincs: A parallel linear constraint solver for molecular simulation},}\ }\href@noop {} {\bibfield  {journal} {\bibinfo  {journal} {Journal of chemical theory and computation}\ }\textbf {\bibinfo {volume} {4}},\ \bibinfo {pages} {116--122} (\bibinfo {year} {2008})}\BibitemShut {NoStop}%
\bibitem [{\citenamefont {Weiss}\ \emph {et~al.}(2011)\citenamefont {Weiss}, \citenamefont {Rullich}, \citenamefont {K{\"o}hler},\ and\ \citenamefont {Frauenheim}}]{weiss2011kinetic}%
  \BibitemOpen
  \bibfield  {author} {\bibinfo {author} {\bibfnamefont {V.~C.}\ \bibnamefont {Weiss}}, \bibinfo {author} {\bibfnamefont {M.}~\bibnamefont {Rullich}}, \bibinfo {author} {\bibfnamefont {C.}~\bibnamefont {K{\"o}hler}}, \ and\ \bibinfo {author} {\bibfnamefont {T.}~\bibnamefont {Frauenheim}},\ }\bibfield  {title} {\enquote {\bibinfo {title} {Kinetic aspects of the thermostatted growth of ice from supercooled water in simulations},}\ }\href@noop {} {\bibfield  {journal} {\bibinfo  {journal} {The Journal of chemical physics}\ }\textbf {\bibinfo {volume} {135}},\ \bibinfo {pages} {034701} (\bibinfo {year} {2011})}\BibitemShut {NoStop}%
\bibitem [{\citenamefont {Espinosa}\ \emph {et~al.}(2016{\natexlab{a}})\citenamefont {Espinosa}, \citenamefont {Navarro}, \citenamefont {Sanz}, \citenamefont {Valeriani},\ and\ \citenamefont {Vega}}]{espinosa2016time}%
  \BibitemOpen
  \bibfield  {author} {\bibinfo {author} {\bibfnamefont {J.}~\bibnamefont {Espinosa}}, \bibinfo {author} {\bibfnamefont {C.}~\bibnamefont {Navarro}}, \bibinfo {author} {\bibfnamefont {E.}~\bibnamefont {Sanz}}, \bibinfo {author} {\bibfnamefont {C.}~\bibnamefont {Valeriani}}, \ and\ \bibinfo {author} {\bibfnamefont {C.}~\bibnamefont {Vega}},\ }\bibfield  {title} {\enquote {\bibinfo {title} {On the time required to freeze water},}\ }\href@noop {} {\bibfield  {journal} {\bibinfo  {journal} {The Journal of chemical physics}\ }\textbf {\bibinfo {volume} {145}},\ \bibinfo {pages} {211922} (\bibinfo {year} {2016}{\natexlab{a}})}\BibitemShut {NoStop}%
\bibitem [{\citenamefont {Matsumoto}, \citenamefont {Yagasaki},\ and\ \citenamefont {Tanaka}(2018)}]{genice}%
  \BibitemOpen
  \bibfield  {author} {\bibinfo {author} {\bibfnamefont {M.}~\bibnamefont {Matsumoto}}, \bibinfo {author} {\bibfnamefont {T.}~\bibnamefont {Yagasaki}}, \ and\ \bibinfo {author} {\bibfnamefont {H.}~\bibnamefont {Tanaka}},\ }\bibfield  {title} {\enquote {\bibinfo {title} {Genice: Hydrogen-disordered ice generator},}\ }\href {\doibase https://doi.org/10.1002/jcc.25077} {\bibfield  {journal} {\bibinfo  {journal} {Journal of Computational Chemistry}\ }\textbf {\bibinfo {volume} {39}},\ \bibinfo {pages} {61--64} (\bibinfo {year} {2018})}\BibitemShut {NoStop}%
\bibitem [{\citenamefont {Vega}, \citenamefont {Abascal},\ and\ \citenamefont {Nezbeda}(2006)}]{vega2006vapor}%
  \BibitemOpen
  \bibfield  {author} {\bibinfo {author} {\bibfnamefont {C.}~\bibnamefont {Vega}}, \bibinfo {author} {\bibfnamefont {J.}~\bibnamefont {Abascal}}, \ and\ \bibinfo {author} {\bibfnamefont {I.}~\bibnamefont {Nezbeda}},\ }\bibfield  {title} {\enquote {\bibinfo {title} {Vapor-liquid equilibria from the triple point up to the critical point for the new generation of tip4p-like models: Tip4p/ew, tip4p/2005, and tip4p/ice},}\ }\href@noop {} {\bibfield  {journal} {\bibinfo  {journal} {The Journal of chemical physics}\ }\textbf {\bibinfo {volume} {125}},\ \bibinfo {pages} {034503} (\bibinfo {year} {2006})}\BibitemShut {NoStop}%
\bibitem [{\citenamefont {Lechner}\ and\ \citenamefont {Dellago}(2008)}]{lechner2008accurate}%
  \BibitemOpen
  \bibfield  {author} {\bibinfo {author} {\bibfnamefont {W.}~\bibnamefont {Lechner}}\ and\ \bibinfo {author} {\bibfnamefont {C.}~\bibnamefont {Dellago}},\ }\bibfield  {title} {\enquote {\bibinfo {title} {Accurate determination of crystal structures based on averaged local bond order parameters},}\ }\href@noop {} {\bibfield  {journal} {\bibinfo  {journal} {The Journal of chemical physics}\ }\textbf {\bibinfo {volume} {129}},\ \bibinfo {pages} {114707} (\bibinfo {year} {2008})}\BibitemShut {NoStop}%
\bibitem [{\citenamefont {Espinosa}\ \emph {et~al.}(2016{\natexlab{b}})\citenamefont {Espinosa}, \citenamefont {Vega}, \citenamefont {Valeriani},\ and\ \citenamefont {Sanz}}]{espinosa2016seeding}%
  \BibitemOpen
  \bibfield  {author} {\bibinfo {author} {\bibfnamefont {J.~R.}\ \bibnamefont {Espinosa}}, \bibinfo {author} {\bibfnamefont {C.}~\bibnamefont {Vega}}, \bibinfo {author} {\bibfnamefont {C.}~\bibnamefont {Valeriani}}, \ and\ \bibinfo {author} {\bibfnamefont {E.}~\bibnamefont {Sanz}},\ }\bibfield  {title} {\enquote {\bibinfo {title} {Seeding approach to crystal nucleation},}\ }\href@noop {} {\bibfield  {journal} {\bibinfo  {journal} {The Journal of chemical physics}\ }\textbf {\bibinfo {volume} {144}} (\bibinfo {year} {2016}{\natexlab{b}})}\BibitemShut {NoStop}%
\bibitem [{\citenamefont {Filion}\ \emph {et~al.}(2010)\citenamefont {Filion}, \citenamefont {Hermes}, \citenamefont {Ni},\ and\ \citenamefont {Dijkstra}}]{filion2010crystal}%
  \BibitemOpen
  \bibfield  {author} {\bibinfo {author} {\bibfnamefont {L.}~\bibnamefont {Filion}}, \bibinfo {author} {\bibfnamefont {M.}~\bibnamefont {Hermes}}, \bibinfo {author} {\bibfnamefont {R.}~\bibnamefont {Ni}}, \ and\ \bibinfo {author} {\bibfnamefont {M.}~\bibnamefont {Dijkstra}},\ }\bibfield  {title} {\enquote {\bibinfo {title} {Crystal nucleation of hard spheres using molecular dynamics, umbrella sampling, and forward flux sampling: A comparison of simulation techniques},}\ }\href@noop {} {\bibfield  {journal} {\bibinfo  {journal} {The Journal of chemical physics}\ }\textbf {\bibinfo {volume} {133}},\ \bibinfo {pages} {244115} (\bibinfo {year} {2010})}\BibitemShut {NoStop}%
\bibitem [{\citenamefont {Yuhara}\ \emph {et~al.}(2015)\citenamefont {Yuhara}, \citenamefont {Barnes}, \citenamefont {Suh}, \citenamefont {Knott}, \citenamefont {Beckham}, \citenamefont {Yasuoka}, \citenamefont {Wu},\ and\ \citenamefont {Sum}}]{Yuhara2015a}%
  \BibitemOpen
  \bibfield  {author} {\bibinfo {author} {\bibfnamefont {D.}~\bibnamefont {Yuhara}}, \bibinfo {author} {\bibfnamefont {B.~C.}\ \bibnamefont {Barnes}}, \bibinfo {author} {\bibfnamefont {D.}~\bibnamefont {Suh}}, \bibinfo {author} {\bibfnamefont {B.~C.}\ \bibnamefont {Knott}}, \bibinfo {author} {\bibfnamefont {G.~T.}\ \bibnamefont {Beckham}}, \bibinfo {author} {\bibfnamefont {K.}~\bibnamefont {Yasuoka}}, \bibinfo {author} {\bibfnamefont {D.}~\bibnamefont {Wu}}, \ and\ \bibinfo {author} {\bibfnamefont {A.~K.}\ \bibnamefont {Sum}},\ }\bibfield  {title} {\enquote {\bibinfo {title} {Nucleation rate analysis of methane hydrate from molecular dynamics simualtions},}\ }\href {\doibase https://doi.org/10.1039/c4fd00219a} {\bibfield  {journal} {\bibinfo  {journal} {Faraday Discuss.}\ }\textbf {\bibinfo {volume} {179}},\ \bibinfo {pages} {463--474} (\bibinfo {year} {2015})}\BibitemShut {NoStop}%
\bibitem [{\citenamefont {Espinosa}, \citenamefont {Vega},\ and\ \citenamefont {Sanz}(2018)}]{espinosa2018homogeneous}%
  \BibitemOpen
  \bibfield  {author} {\bibinfo {author} {\bibfnamefont {J.~R.}\ \bibnamefont {Espinosa}}, \bibinfo {author} {\bibfnamefont {C.}~\bibnamefont {Vega}}, \ and\ \bibinfo {author} {\bibfnamefont {E.}~\bibnamefont {Sanz}},\ }\bibfield  {title} {\enquote {\bibinfo {title} {Homogeneous ice nucleation rate in water droplets},}\ }\href@noop {} {\bibfield  {journal} {\bibinfo  {journal} {The Journal of Physical Chemistry C}\ }\textbf {\bibinfo {volume} {122}},\ \bibinfo {pages} {22892--22896} (\bibinfo {year} {2018})}\BibitemShut {NoStop}%
\bibitem [{\citenamefont {Gibbs}(1876)}]{gibbsCNT1}%
  \BibitemOpen
  \bibfield  {author} {\bibinfo {author} {\bibfnamefont {J.~W.}\ \bibnamefont {Gibbs}},\ }\bibfield  {title} {\enquote {\bibinfo {title} {On the equilibrium of heterogeneous substances},}\ }\href@noop {} {\bibfield  {journal} {\bibinfo  {journal} {Trans. Connect. Acad. Sci.}\ }\textbf {\bibinfo {volume} {3}},\ \bibinfo {pages} {108--248} (\bibinfo {year} {1876})}\BibitemShut {NoStop}%
\bibitem [{\citenamefont {Gibbs}(1878)}]{gibbsCNT2}%
  \BibitemOpen
  \bibfield  {author} {\bibinfo {author} {\bibfnamefont {J.~W.}\ \bibnamefont {Gibbs}},\ }\bibfield  {title} {\enquote {\bibinfo {title} {On the equilibrium of heterogeneous substances},}\ }\href@noop {} {\bibfield  {journal} {\bibinfo  {journal} {Trans. Connect. Acad. Sci.}\ }\textbf {\bibinfo {volume} {16}},\ \bibinfo {pages} {343--524} (\bibinfo {year} {1878})}\BibitemShut {NoStop}%
\bibitem [{\citenamefont {Volmer}\ and\ \citenamefont {Weber}(1926)}]{volmerweber1926}%
  \BibitemOpen
  \bibfield  {author} {\bibinfo {author} {\bibfnamefont {M.}~\bibnamefont {Volmer}}\ and\ \bibinfo {author} {\bibfnamefont {A.}~\bibnamefont {Weber}},\ }\bibfield  {title} {\enquote {\bibinfo {title} {Keimbildung in ubersattigten gebilden},}\ }\href@noop {} {\bibfield  {journal} {\bibinfo  {journal} {Z. Phys. Chem.}\ }\textbf {\bibinfo {volume} {119}},\ \bibinfo {pages} {277} (\bibinfo {year} {1926})}\BibitemShut {NoStop}%
\bibitem [{\citenamefont {Becker}\ and\ \citenamefont {Doring}(1935)}]{becker-doring}%
  \BibitemOpen
  \bibfield  {author} {\bibinfo {author} {\bibfnamefont {R.}~\bibnamefont {Becker}}\ and\ \bibinfo {author} {\bibfnamefont {W.}~\bibnamefont {Doring}},\ }\bibfield  {title} {\enquote {\bibinfo {title} {Kinetische behandlung der keimbildung in ubersattigten dampfen},}\ }\href@noop {} {\bibfield  {journal} {\bibinfo  {journal} {Ann. Phys.}\ }\textbf {\bibinfo {volume} {416}},\ \bibinfo {pages} {719--752} (\bibinfo {year} {1935})}\BibitemShut {NoStop}%
\bibitem [{\citenamefont {Espinosa}\ \emph {et~al.}(2014)\citenamefont {Espinosa}, \citenamefont {Sanz}, \citenamefont {Valeriani},\ and\ \citenamefont {Vega}}]{espinosa2014homogeneous}%
  \BibitemOpen
  \bibfield  {author} {\bibinfo {author} {\bibfnamefont {J.}~\bibnamefont {Espinosa}}, \bibinfo {author} {\bibfnamefont {E.}~\bibnamefont {Sanz}}, \bibinfo {author} {\bibfnamefont {C.}~\bibnamefont {Valeriani}}, \ and\ \bibinfo {author} {\bibfnamefont {C.}~\bibnamefont {Vega}},\ }\bibfield  {title} {\enquote {\bibinfo {title} {Homogeneous ice nucleation evaluated for several water models},}\ }\href@noop {} {\bibfield  {journal} {\bibinfo  {journal} {The Journal of chemical physics}\ }\textbf {\bibinfo {volume} {141}},\ \bibinfo {pages} {18C529} (\bibinfo {year} {2014})}\BibitemShut {NoStop}%
\bibitem [{\citenamefont {Volmer}(1929)}]{volmer1929keimbildung}%
  \BibitemOpen
  \bibfield  {author} {\bibinfo {author} {\bibfnamefont {M.}~\bibnamefont {Volmer}},\ }\bibfield  {title} {\enquote {\bibinfo {title} {{\"U}ber keimbildung und keimwirkung als spezialf{\"a}lle der heterogenen katalyse},}\ }\href@noop {} {\bibfield  {journal} {\bibinfo  {journal} {Zeitschrift f{\"u}r Elektrochemie und angewandte physikalische Chemie}\ }\textbf {\bibinfo {volume} {35}},\ \bibinfo {pages} {555--561} (\bibinfo {year} {1929})}\BibitemShut {NoStop}%
\bibitem [{\citenamefont {Camarillo}\ \emph {et~al.}(2025)\citenamefont {Camarillo}, \citenamefont {Oller-Iscar}, \citenamefont {M~Conde}, \citenamefont {Ram{\'\i}rez},\ and\ \citenamefont {Sanz}}]{camarillo2025heterogenous}%
  \BibitemOpen
  \bibfield  {author} {\bibinfo {author} {\bibfnamefont {M.}~\bibnamefont {Camarillo}}, \bibinfo {author} {\bibfnamefont {J.}~\bibnamefont {Oller-Iscar}}, \bibinfo {author} {\bibfnamefont {M.}~\bibnamefont {M~Conde}}, \bibinfo {author} {\bibfnamefont {J.}~\bibnamefont {Ram{\'\i}rez}}, \ and\ \bibinfo {author} {\bibfnamefont {E.}~\bibnamefont {Sanz}},\ }\bibfield  {title} {\enquote {\bibinfo {title} {Heterogeneous ice nucleation on model substrates},}\ }\href@noop {} {\bibfield  {journal} {\bibinfo  {journal} {The Journal of Chemical Physics}\ }\textbf {\bibinfo {volume} {163}},\ \bibinfo {pages} {154504} (\bibinfo {year} {2025})}\BibitemShut {NoStop}%
\bibitem [{\citenamefont {Young}(1805)}]{young1805iii}%
  \BibitemOpen
  \bibfield  {author} {\bibinfo {author} {\bibfnamefont {T.}~\bibnamefont {Young}},\ }\bibfield  {title} {\enquote {\bibinfo {title} {{III}. an essay on the cohesion of fluids},}\ }\href@noop {} {\bibfield  {journal} {\bibinfo  {journal} {Philosophical transactions of the royal society of London}\ ,\ \bibinfo {pages} {65--87}} (\bibinfo {year} {1805})}\BibitemShut {NoStop}%
\bibitem [{\citenamefont {Dellago}, \citenamefont {Bolhuis},\ and\ \citenamefont {Geissler}(2002)}]{dellago2002transition}%
  \BibitemOpen
  \bibfield  {author} {\bibinfo {author} {\bibfnamefont {C.}~\bibnamefont {Dellago}}, \bibinfo {author} {\bibfnamefont {P.~G.}\ \bibnamefont {Bolhuis}}, \ and\ \bibinfo {author} {\bibfnamefont {P.~L.}\ \bibnamefont {Geissler}},\ }\bibfield  {title} {\enquote {\bibinfo {title} {Transition path sampling},}\ }\href@noop {} {\bibfield  {journal} {\bibinfo  {journal} {Advances in chemical physics}\ }\textbf {\bibinfo {volume} {123}},\ \bibinfo {pages} {1--78} (\bibinfo {year} {2002})}\BibitemShut {NoStop}%
\bibitem [{\citenamefont {Moroni}, \citenamefont {Ten~Wolde},\ and\ \citenamefont {Bolhuis}(2005)}]{moroni2005interplay}%
  \BibitemOpen
  \bibfield  {author} {\bibinfo {author} {\bibfnamefont {D.}~\bibnamefont {Moroni}}, \bibinfo {author} {\bibfnamefont {P.~R.}\ \bibnamefont {Ten~Wolde}}, \ and\ \bibinfo {author} {\bibfnamefont {P.~G.}\ \bibnamefont {Bolhuis}},\ }\bibfield  {title} {\enquote {\bibinfo {title} {Interplay between structure and size in a critical crystal nucleus},}\ }\href@noop {} {\bibfield  {journal} {\bibinfo  {journal} {Physical review letters}\ }\textbf {\bibinfo {volume} {94}},\ \bibinfo {pages} {235703} (\bibinfo {year} {2005})}\BibitemShut {NoStop}%
\bibitem [{\citenamefont {Moore}\ and\ \citenamefont {Molinero}(2011)}]{moore2011structural}%
  \BibitemOpen
  \bibfield  {author} {\bibinfo {author} {\bibfnamefont {E.~B.}\ \bibnamefont {Moore}}\ and\ \bibinfo {author} {\bibfnamefont {V.}~\bibnamefont {Molinero}},\ }\bibfield  {title} {\enquote {\bibinfo {title} {Structural transformation in supercooled water controls the crystallization rate of ice},}\ }\href@noop {} {\bibfield  {journal} {\bibinfo  {journal} {Nature}\ }\textbf {\bibinfo {volume} {479}},\ \bibinfo {pages} {506--508} (\bibinfo {year} {2011})}\BibitemShut {NoStop}%
\bibitem [{\citenamefont {Hudait}\ \emph {et~al.}(2016)\citenamefont {Hudait}, \citenamefont {Qiu}, \citenamefont {Lupi},\ and\ \citenamefont {Molinero}}]{hudait2016free}%
  \BibitemOpen
  \bibfield  {author} {\bibinfo {author} {\bibfnamefont {A.}~\bibnamefont {Hudait}}, \bibinfo {author} {\bibfnamefont {S.}~\bibnamefont {Qiu}}, \bibinfo {author} {\bibfnamefont {L.}~\bibnamefont {Lupi}}, \ and\ \bibinfo {author} {\bibfnamefont {V.}~\bibnamefont {Molinero}},\ }\bibfield  {title} {\enquote {\bibinfo {title} {Free energy contributions and structural characterization of stacking disordered ices},}\ }\href@noop {} {\bibfield  {journal} {\bibinfo  {journal} {Physical Chemistry Chemical Physics}\ }\textbf {\bibinfo {volume} {18}},\ \bibinfo {pages} {9544--9553} (\bibinfo {year} {2016})}\BibitemShut {NoStop}%
\end{thebibliography}
\end{document}